\documentclass[letterpaper]{ar-1col}
\usepackage{showyourwork}
\usepackage[letterpaper]{geometry}

\usepackage{natbib}
\usepackage{amsmath}
\usepackage{color}
\usepackage{hyperref}
\hypersetup{hidelinks}

\usepackage{graphbox}

\usepackage{url}

\jname{Annu. Rev. Astron. Astrophys.}
\jvol{AA}
\jyear{2022}
\doi{10.1146/TBD}

\newcommand{\ydata}{\ensuremath{\boldsymbol{y}}}
\newcommand{\hyperparams}{\ensuremath{\boldsymbol{\phi}}}
\newcommand{\meanparams}{\ensuremath{\boldsymbol{\theta}}}
\newcommand{\dt}{\ensuremath{\tau}}
\newcommand{\amplitude}{\ensuremath{\alpha}}
\newcommand{\lengthscale}{\ensuremath{\lambda}}

\DeclareMathOperator*{\argmax}{arg\,max}

\newcommand{\project}[1]{\textsf{#1}}

\begin{document}

\markboth{Aigrain \& Foreman-Mackey}{GPR for Astronomy}

\title{Gaussian Process regression for astronomical time-series}

\author{Suzanne Aigrain,$^1$ and Daniel Foreman-Mackey$^2$
  \affil{$^1$Department of Physics, University of Oxford, Oxford, UK, OX1 3RH; email: suzanne.aigrain@physics.ox.ac.uk}
  \affil{$^2$Center for Computational Astrophysics, Flatiron Institute, New York, USA, NY 10010; email: dforeman-mackey@flatironinstitute.org}}

\begin{abstract}
The last two decades have seen a major expansion in the availability, size, and precision of time-domain datasets in astronomy. Owing to their unique combination of flexibility, mathematical simplicity and comparative robustness, Gaussian Processes (GPs) have emerged recently as the solution of choice to model stochastic signals in such datasets. In this review we provide a brief introduction to the emergence of GPs in astronomy, present the underlying mathematical theory, and give practical advice considering the key modelling choices involved in GP regression. We then review applications of GPs to time-domain datasets in the astrophysical literature so far, from exoplanets to active galactic nuclei, showcasing the power and flexibility of the method. We provide worked examples using simulated data, with links to the source code, discuss the problem of computational cost and scalability, and give a snapshot of the current ecosystem of open source GP software packages.
Driven by further algorithmic and conceptual advances, we expect that GPs will continue to be an important tool for robust and interpretable time domain astronomy for many years to come.
\end{abstract}

\maketitle

\tableofcontents

\section{INTRODUCTION}
\label{sec:intro}

Gaussian Processes (GPs) are a powerful class of statistical models, which allow us to define a probability distribution over random functions. Rather than write down an explicit mathematical formula for the function from which some observations are generated, we model the covariance between pairs of samples from the process, using our physical domain knowledge and/or available data to guide our modelling.
While this may seem abstract, GPs have a wide range of applications from modelling of stochastic physical processes, to high dimensional interpolation and smoothing.
\begin{armarginnote}[]
  \entry{GP}{Gaussian Process}
  \entry{GPR}{Gaussian Process Regression}
\end{armarginnote}
In particular, GP Regression (GPR) has become increasingly popular in the astronomical community over the last decade. Part of the reason for this uptake is the growing availability and importance of time-domain datasets in astronomy. These systematically contain non-trivial random or unknown signals, whether astrophysical or instrumental, that need to be modelled. In many cases, these are nuisance signals, which we need to marginalise over in order to detect or measure other signals robustly. Sometimes, we are interested in the stochastic behaviour itself, and want to infer its characteristics or predict its behaviour. GPR offers a compelling solution: statistically principled, naturally Bayesian, and extremely flexible, yet mathematically simple. To new users, however, the lack of an explicit functional form for the model can make GPR can seem a little arcane. Furthermore, the computational cost of the method, which naively scales cubically with the dataset size, can be an obstacle. These factors initially impeded its dissemination in the astronomical community, but have been largely overcome in recent years thanks to the availability of user-friendly, computationally optimised software packages.

In this review, we aim to give the reader a concise introduction to the basic theoretical framework of GPR, and to illustrate the range, strengths and limitations of the method by discussing some of its applications to time-domain datasets in astronomy so far. We made no attempt to provide a complete theoretical description or to discuss the relevant literature exhaustively: there are excellent textbooks which already do the former, while the latter would have been impossible in the space available. Instead, our goal was to provide an accessible introduction that might give interested readers starting points for further reading, and help them determine whether GPR might be useful for their own datasets. With this in mind, we took pains to require minimal prior knowledge, to provide practical advice regarding the modelling choices one needs to make when using GPR, and to review relevant open-source software.


As a technical note, this manuscript was prepared using the \project{showyourwork} package\footnote{\url{https://show-your.work}} and the source code used to generate each figure is available in a public \project{GitHub} repository\footnote{\url{https://github.com/dfm/araa-gps}}.
To see the specific version of the \project{Jupyter} notebook, that was executed to generate each figure, click on the icon next to the figure caption.

\subsection{Brief history}

An early use of GPR, was for spatial interpolation in geophysics \citep{kriging}, and GPR has since been adopted or re-invented in a wide range of other application domains. GPs were used in simulations in a wide range of astronomical subfields \citep[see e.g.][]{1980asfr.symp..159B,1988JGR....9311569C,1997ApJ...483L...1P}, but early mentions of GPs for modelling astronomical datasets \citep[see e.g.][]{1976MitAG..38..192D,1978A&A....70..777V,1991MGeo...16..313J} received limited attention.

\begin{figure}[ht]
  \centering
  \script{literature.ipynb}
  \includegraphics[width=0.5\linewidth]{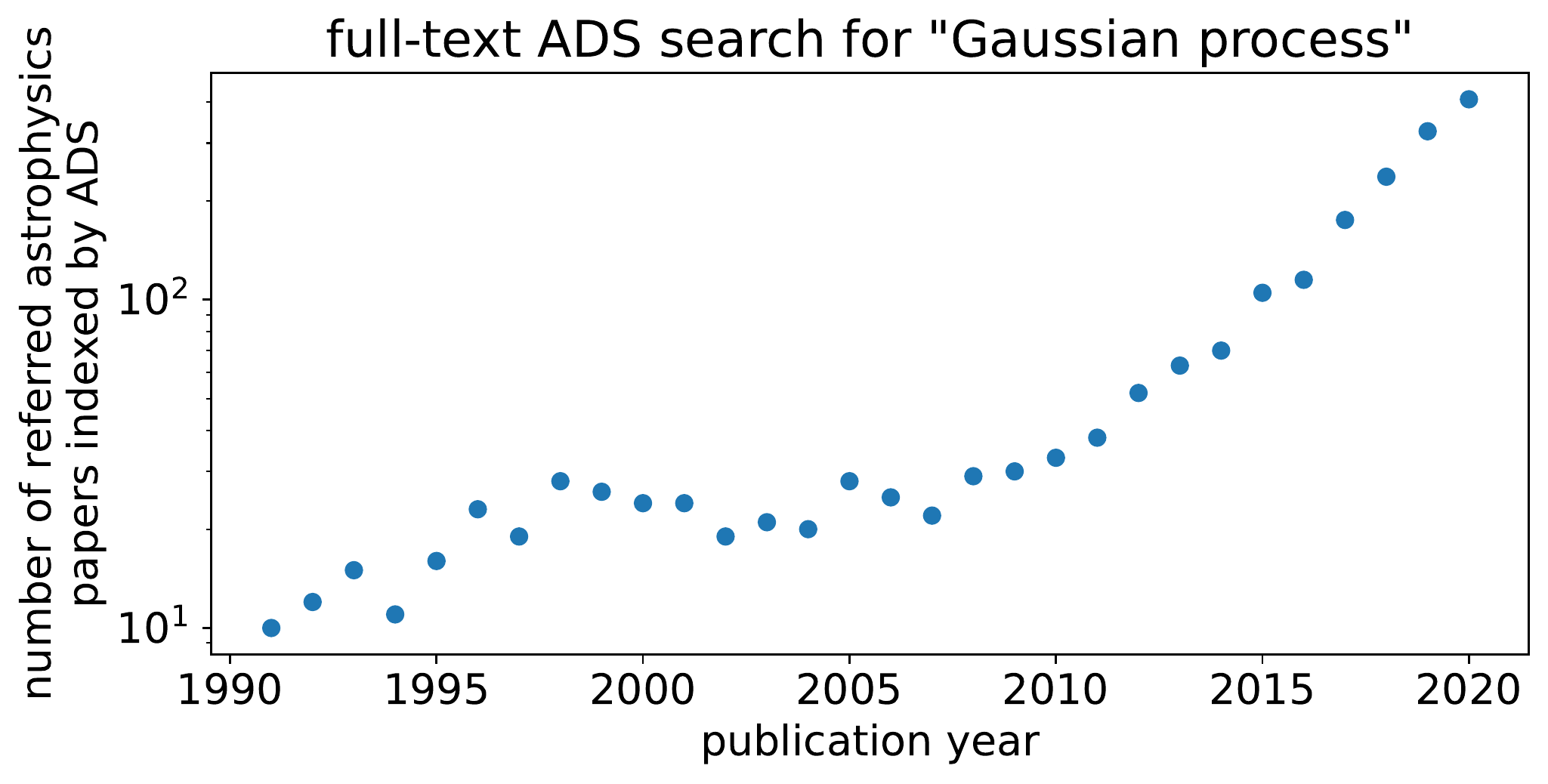}
  \caption{The number of referred publications in the astronomy and astrophysics literature that include the text ``Gaussian process'' as indexed by the NASA Astrophysics Data System (ADS). }
  \label{fig:literature}
\end{figure}

Perhaps the earliest use of GPR in the refereed astronomical literature that will be familiar to a modern reader was published by \citet{prh92a} in the context of quasar variability, and for a long time this remained its main application domain in astronomy. GPR then gradually appeared in other areas, starting with photometric redshift estimation \citep{2006ApJ...647..102W}, then exoplanet transit observations \citep{2009ApJ...704...51C,2012MNRAS.419.2683G} and radial velocity planet searches \citep{2012MNRAS.419.3147A,2014MNRAS.443.2517H}. Nonetheless, GPR remained relatively niche and few astronomers had heard of it until a few years ago. To illustrate this point, we searched on the NASA Astrophysics Data System (ADS) for articles published in refereed astronomy and astrophysics journals with the words ``Gaussian Process'' in the full text of the article \autoref{fig:literature}.
After an increase in popularity throughout the 1990s, the use of GPs in astrophysics remained fairly constant around $\sim20$ publications per year until 2010.
Since 2010, the popularity of GPs has grown significantly, and in 2021, more than 500 refereed papers referencing GPs were published in the astrophysics literature. A snapshot of the areas where GPs are currently used in astronomy can be obtained readily from the keywords of those papers published in the last two years (using the NASA ADS analytics tools). This shows that approximately a third of these are cosmology papers, where GPs are primarily used as emulators; this falls outside the scope of this review since it doesn't involve time-series data. The second and largest grouping covers exoplanets and, by extension, stellar variability (since most exoplanet discoveries rely on indirect detection of the planetary signals in observations of their host stars). Finally, the remaining 30\% or so of the papers relate to other kinds of astrophysical variability including active galactic nuclei, compact objects, transients and gravitational waves.

A number of important factors have contributed to recent the democratisation of GPR across a wide range of scientific disciplines, including the publication of a dedicated textbook \citep{gpml}, as well as the availability of cheap computing power and user-friendly, open-source GPR software. In astronomy, specifically, packages such as \project{george} \citep{george} and \project{celerite} \citep{celerite}, which were developed by astronomers working with applied mathematicians, have been instrumental in broadening the user base for GPR. An additional factor has also been at play in this application domain: the rise of time domain surveys. Correlated noise in time-domain observations is a direct and unavoidable consequence of causation, and hence ubiquitous. Adequately modelling this correlated noise is vital when searching for faint signals, for example from exoplanets. Astrophysical sources, from accretion disks on all scales to magnetically active stars or cloudy brown dwarfs, also display complex, intrinsically or apparently stochastic behaviour, for which adequate modelling strategies are required. GPR is a natural choice to tackle these challenges. Rather than attempting to cover all applications of GPs in astrophysics, which would not be feasible in the space available, we have therefore opted to focus on its application to time-domain datasets in astronomy.

\subsection{Motivating examples}
\label{sec:sim_examples}

By way of motivation, before delving any further into the methodology, we have produced two illustrative examples that highlight some typical use cases for GPR in astrophysics.
These examples are not meant to comprehensively summarize the space of use cases.
Similarly, the goal of this section is not to formally compare the performance of GPR to other methods with the same goals.
Instead, these examples are meant to identify some common, but qualitatively different applications.
In both examples, we use simulated datasets since it is useful to know the ``ground truth'' to validate performance.

The first example is a re-implementation of one of the earliest uses of GPR in time-domain astronomy \citep{prh92a}, using modern language and techniques.
In this example, we measure the time delay of a lensed quasar, using a GP as a flexible, non-parametric model for the latent (unobserved) variability of the unlensed quasar system.

The second example demonstrates the use of GPR to account for stellar variability in the light curve of a transiting exoplanet, when inferring its parameters.
In this case, the parameters of interest are the parameters of the mean model, and the GP is a nuisance model, and our goal is to propagate uncertainty introduced by the stellar variability to our constraints on the physical parameters of interest.

These examples---and all the examples throughout this review---are implemented using \textsf{tinygp}\footnote{\url{https://tinygp.readthedocs.io}} \citep{tinygp}, a \project{Python} library for GPR built on top of the \project{JAX}\footnote{\url{https://jax.readthedocs.io}} library for numerical computing \citep{jax}.
Here and throughout, the probabilistic models are implemented using the \project{NumPyro}\footnote{\url{https://num.pyro.ai}} library \citep{numpyro} and the Markov chain Monte Carlo (MCMC) inference is performed using the No U-Turn Sampling (NUTS) algorithm \citep{Hoffman:2014}.

\subsubsection{Example 1: The time delay of a gravitational lensed quasar}
\label{sec:quasar}

In this example, we re-visit the method developed by \citet{prh92a} to measure the time delay of the gravitationally lensed quasar 0957+561, one of the earliest applications of GPR for time domain astronomy. 
The underlying model here is that the unobserved latent variability of the source is modelled using a GP, in this case we use a Mat\'ern-3/2 covariance function as discussed and defined in \autoref{sec:kernels}.
The images sample this time series at lagged times and with different mean magnitudes and variability amplitudes. 

\begin{figure}[ht]
  \centering
  \script{quasar.ipynb}
  \begin{minipage}[t]{0.5\linewidth}
    \includegraphics[width=\linewidth,vshift=1cm]{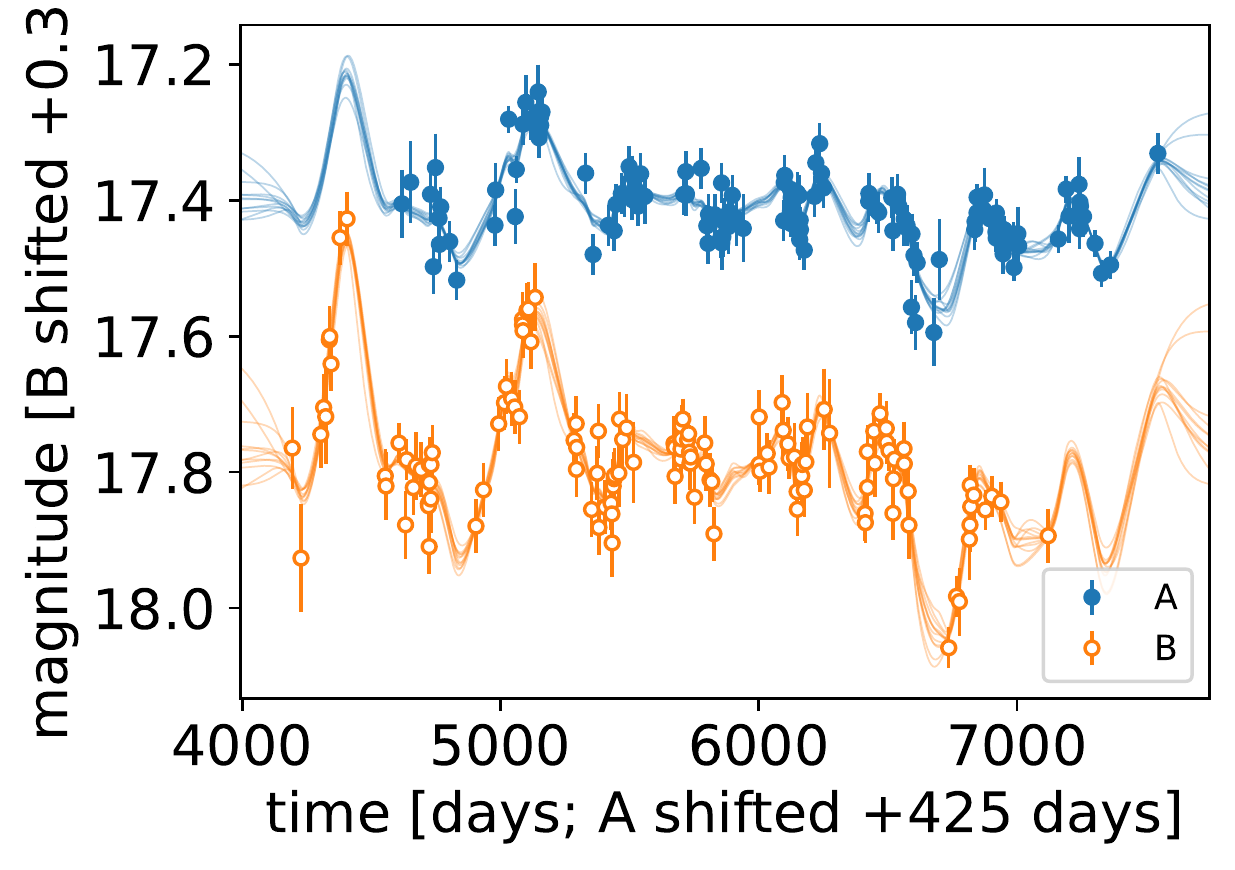}
  \end{minipage} \hfill
  \begin{minipage}[t]{0.44\linewidth}
    \includegraphics[width=\linewidth]{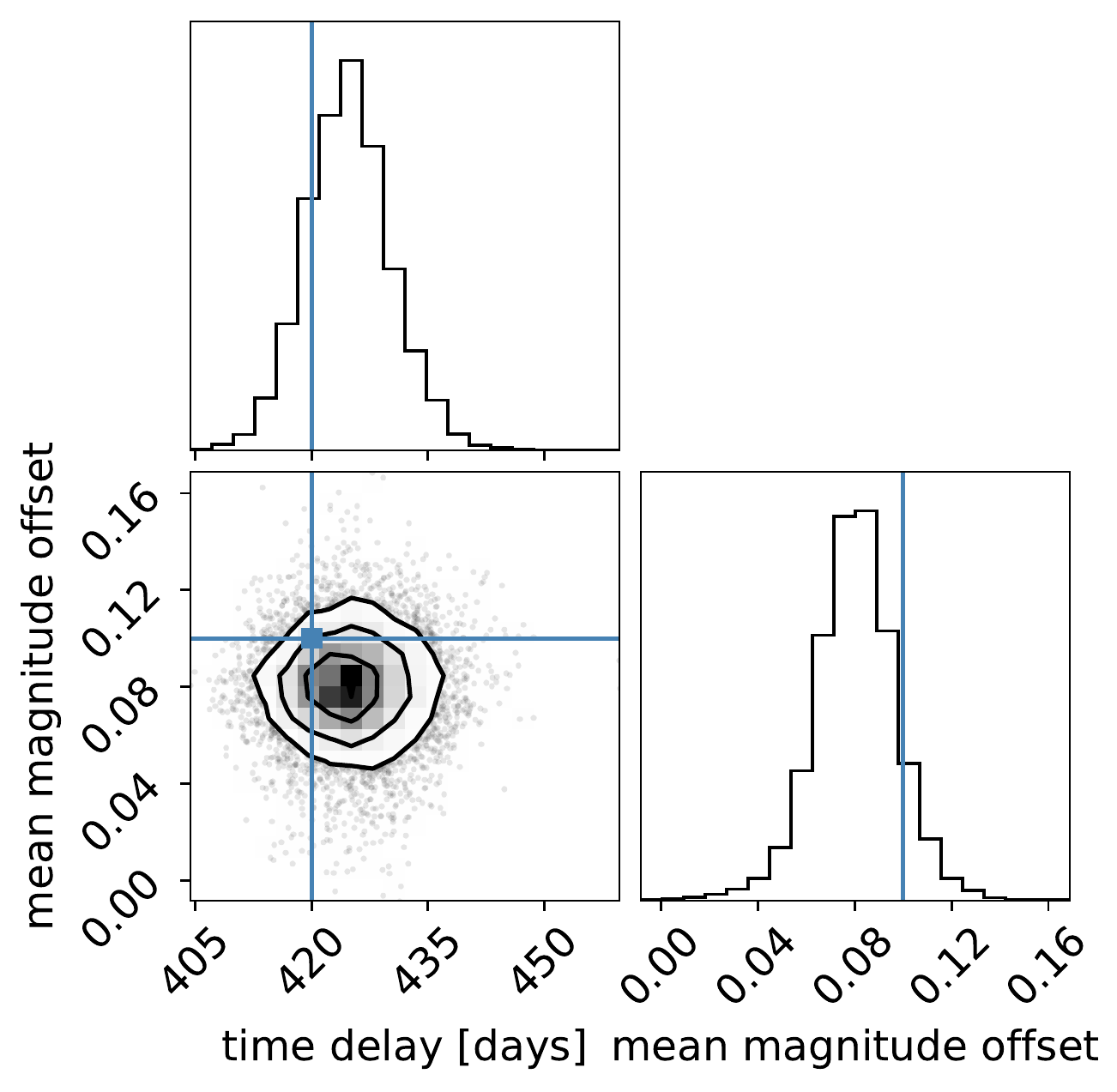}
  \end{minipage}
  \caption{The results of fitting the simulated light curves of two images of a lensed quasar with a time delay between images.
  \emph{Left:} The simulated light curves for each image with an arbitrary magnitude offset applied to image B, and the median of posterior time delay applied to image A, bringing the light curves into a common frame.
  The data are plotted as points with error bars, and over-plotted on these data are posterior predictive samples of each images' predicted variability.
  \emph{Right:} The posterior constraints on the time delay and mean magnitude offset between the two images, obtained using Markov chain Monte Carlo (MCMC) to fit the simulated data shown in the left panel of this figure.
  The true values of these parameters that were used to simulate the data are over-plotted as blue lines.}
  \label{fig:quasar}
\end{figure}

Under this assumed model, we simulate a pair of light curves with the same cadence and uncertainties as the dataset from \citet{1989A&A...215....1V} that was analysed by \citet{prh92a}. 
In this simulation, the parameters of the covariance model, the mean magnitudes, and the time delay are all set to known values, designed to produce qualitatively similar features to the dataset from \citet{1989A&A...215....1V}.
The simulated light curves are plotted in the left panel of \autoref{fig:quasar}.

Using these simulated data, we fit a GP model using MCMC, varying the time delay, the mean magnitude of each image, the variability amplitude of each image, and the timescale of the covariance.
The results of this inference are shown in \autoref{fig:quasar}.
In the left panel we show the simulated data with the median of posterior time delay applied to image A, and an arbitrary magnitude offset applied to image B for plotting purposes.
Over-plotted on these data are 12 posterior samples of the GP model predictions for the noise-free photometry for each image.
This figure captures how a GP can be used to flexibly capture a stochastically variable process under certain smoothness constraints, and how the uncertainties on the interpolated and extrapolated predictions increase away from the observed data.

The right panel of \autoref{fig:quasar} shows the posterior constraints for two of the key parameters of the model: the time delay, and the mean magnitude difference between images.
Since these are simulated data, we know the true values of these parameters, and these true values are over-plotted on \autoref{fig:quasar}b, demonstrating that our method reliably recovers the expected result.

It is worth noting that the determination of the time-delay for 0957+561 attracted considerable controversy at the time \citep[see][and references therein]{1995ApJ...455L...5K} and that later observations \citep{1997ApJ...482...75K} favour a significantly shorter time-delay than that obtained by \citet{prh92a}. Whatever the statistical method used, 
the likelihood function for translation problems of this type can be highly multi-modal on large scales, making identification of the global optimum challenging. Furthermore, the assumption that one quasar image's light curve is merely a scaled and shifted version of the other is probably too simplistic. This is another reason why we used simulated data in this example, rather than work with the  dataset originally analysed by \citet{prh92a}.

\subsubsection{Example 2: Fitting an exoplanet transit with stellar variability}
\label{sec:transit}

In this second example, we demonstrate another common application of GPR in astrophysics, as a flexible model for nuisances and correlated noise, where we want to correctly capture uncertainty introduced by this noise model into our constraints on the parameters of interest.
To this end, we simulate the light curve of a transiting exoplanet and aim to infer the physical parameters of the system, taking correlated noise into account.
The correlated noise in transit light curves is typically caused by the variability of the host star and by instrumental effects like focus or pointing changes.
It has been demonstrated that neglecting to account for these variations can cause significant errors when inferring the properties of the planet \citep{2006MNRAS.373..231P,2007A&A...472L..13G}.
This correlated noise can often be well modelled by a GP, and the use of GPs for transit modelling has been a fruitful area of research in the astrophysics literature (see \autoref{sec:transit_fit} for a more detailed discussion).

\begin{figure}[ht]
  \centering
  \script{transit.ipynb}
  \begin{minipage}[t]{0.5\linewidth}
    \includegraphics[width=\linewidth,vshift=1cm]{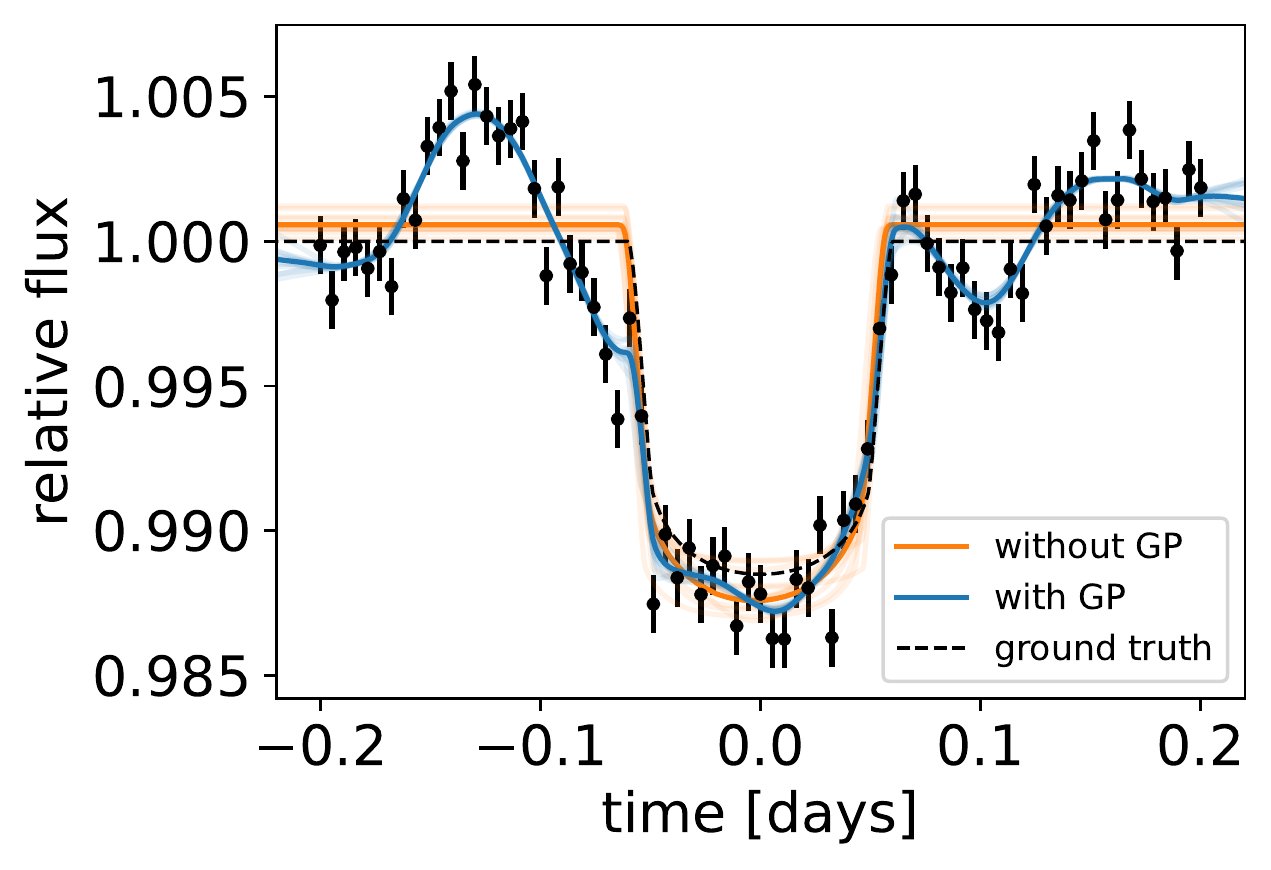}
  \end{minipage} \hfill
  \begin{minipage}[t]{0.44\linewidth}
    \includegraphics[width=\linewidth]{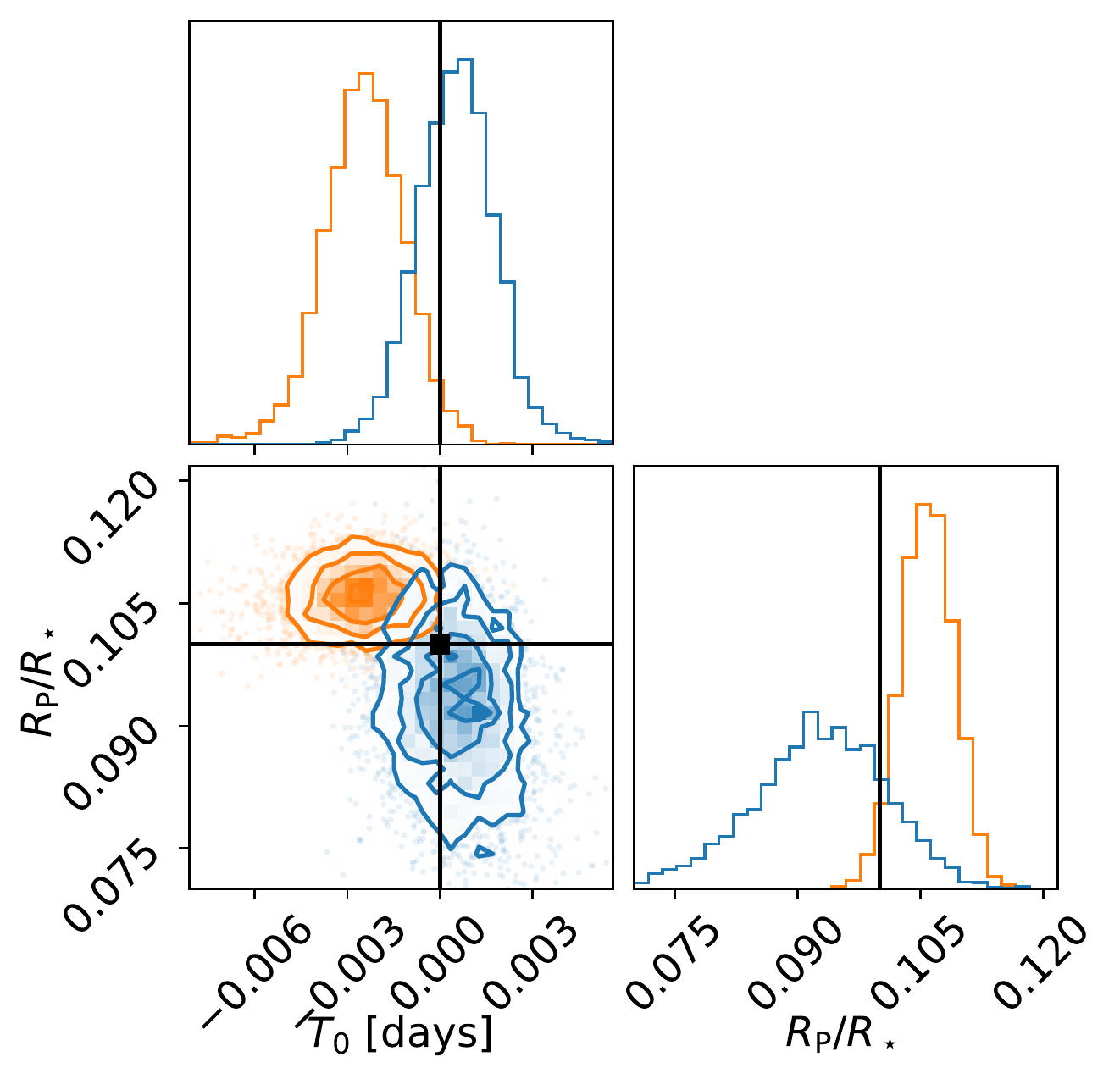}
  \end{minipage}
  \caption{The results of fitting the simulated light curve of a transiting exoplanet.
  \emph{Left:} The simulated dataset plotted as points with error bars.
  The inferred model when correlated noise is (blue) or is not (orange) taken into account are over-plotted on the data.
  \emph{Right:} The posterior constraints on the time of transit $T_0$ and the planet-to-star radius ratio $R_\mathrm{p}/R_\star$.
  The colours in this figure match those in the left panel, and the true values of these parameters are indicated with black lines.
  The results when neglecting correlated noise (orange) are significantly inconsistent with the true value, but when the correlated noise is modelled using a GP (blue), the correct parameters are recovered, albeit with larger uncertainty.}
  \label{fig:transit}
\end{figure}

For this example, the transit was simulated with known physical properties such as the planet-to-star radius ratio, the impact parameter, and quadratic limb darkening parameters \citep{2020AJ....159..123A}.
We then add correlated and white noise by sampling from a GP model with a Mat\'ern-3/2 covariance function as defined in \autoref{sec:kernels}, with a known amplitude, timescale, and white noise amplitude.
These simulated data are shown in the left panel of \autoref{fig:transit}.

We then used the same framework to model the simulated dataset, first fitting only for an excess variance or ``jitter'' term (a method commonly used to account for model mis-specification) and then accounting for the correlated nature of the explicitly using a GP.
For the transit model, we fit for the planet-to-star radius ratio $R_{\rm p}/R_\star$, the mid-transit time $T_0$, the out-of-transit flux $f_0$, and the limb darkening parameters $u_1,\,u_2$.
We then marginalise over either the variance of the jitter term, or the amplitude $\amplitude$ and timescale $\lengthscale$ of the GP noise model.
The posterior constraints on the time of transit and the planet-to-star ratio are shown in the right panel of \autoref{fig:transit}, with the true values of these parameters over-plotted as a black line.
When the correlated noise is absorbed into a jitter term (plotted in orange in \autoref{fig:transit}), the inferred parameters are significantly inconsistent with the truth.
Taking the correlated noise into account (blue in \autoref{fig:transit}) increases the uncertainties on the inferred parameters, but also shifts the results to recover the true parameters.


\subsection{Overview of this review}

The remainder of this review is structured as follows. \autoref{sec:basics} provides a brief, accessible introduction to the basic theory of GPR, and \autoref{sec:choices} discusses the key modelling choices one needs to make when using a GP to model data, with practical advice for how to make these choices. \autoref{sec:uses} gives an overview of applications of GPR to time-domain astronomical datasets to date, from stars and exoplanets to active galactic nuclei, pulsars and gravitational waves. In \autoref{sec:challenges} we discuss some challenges of using GPR in practice, and how they may be overcome, while \autoref{sec:open} reviews the ecosystem of open-source GPR software available today. Finally, we summarise the main take-away points and outline future directions in \autoref{sec:concl}.

\section{BASICS OF GAUSSIAN PROCESS REGRESSION}
\label{sec:basics}

So far we have discussed how GPR has become widely used in astronomy and touched on some of their pros and cons, but we have not explicitly defined what a GP is or explained how GPR works. This section gives a brief introduction to the theory of GPR, written in a way that we hope will be intelligible to most astronomers. A much more in-depth treatment is provided by \citet{gpml}, which to this day remains the main reference textbook on GPs.

\subsection{Formal definition}
\label{sec:def}

A GP is a type of \textit{stochastic process} based on the Gaussian probability distribution. A probability distribution describes a random variable with a finite number of dimensions. A stochastic process extends this concept to an infinite number of dimensions, allowing us to define a probability distribution over functions. Just what do we mean by ``extending to an infinite number of dimensions''? Well, this can be a little problematic mathematically, but we need not worry about it, because in practice we only ever deal with finite samples from the stochastic process.

The formal definition of a GP is that the joint probability distribution over any finite sample $\mathbf{y} = \{y_i\}_{i=1,\ldots,N}$ from the GP is a multi-variate Gaussian:
\begin{equation}
\label{eq:gp_def}
    p(\mathbf{y}) = \mathcal{N}(\mathbf{m}, \mathbf{K}),
\end{equation}
where $\mathbf{m}$ is the \emph{mean vector} and $\mathbf{K}$ the \emph{covariance matrix}.

The elements of the mean vector and covariance matrix are given by the \textit{mean function} $m$ and the \textit{covariance function} $k$, respectively:
\begin{equation}
    m_{i} = m(\boldsymbol{x}_i, \meanparams),
\end{equation}
\begin{equation}
    K_{ij} = k(\boldsymbol{x}_i,\boldsymbol{x}_j, \hyperparams).
\end{equation}
where $\boldsymbol{x}_i$ is the set of inputs (independent variables) corresponding to the $i^{\rm th}$ sample. For time-series data, the inputs usually include, but are not necessarily restricted to, the time $t_i$. The covariance function, also known as the \emph{kernel function}, is the fundamental ingredient of a GP model, and considerable care must be taken to select it adequately (or to test different possibilities). Sometimes the mean function is assumed to be constant, or even zero, everywhere; this is often done in the wider GPR literature to keep derivations uncluttered. However, for many astrophysical applications where a GP is used to model a nuisance signal, the mean function contains the signal of interest and is important.


The parameters $\meanparams$ and $\hyperparams$ of the mean and covariance function are known as the \textit{hyperparameters} of the GP. Strictly speaking, the \textit{parameters} of the GP are the (infinitely many) unknown functions that share the specified mean vector and covariance matrix and could have given rise to the observations. However, these parameters are always marginalised over: we never explicitly deal with the individual functions, except when drawing samples for illustrative purposes (as we did in Figures~\ref{fig:quasar} and \ref{fig:transit}). GPs are therefore a type of \textit{Hierarchical Bayesian Model} (HBM). 
We never observe the unknown function that gave rise to the data directly, but infer a probability distribution for it from our noisy observations.

Although it is possible to construct and use stochastic process models based on other distributions, GPs are by far the most popular, for two main reasons. The first is Central Limit theorem: it implies that the assumption of Gaussianity is often at least approximately correct. The second is that Gaussian distributions obey simple mathematical identities for marginalisation and conditioning, that enable inference with GPs (the process of marginalising over the individual functions) to be performed \textit{analytically}, with very simple linear algebra. It is this analytic marginalisation property that sets GPR apart from other forms of HBM. 

\subsection{From least-squares to GPR}
\label{sec:from_lsq}

The formal definition given in \autoref{eq:gp_def} does not necessarily provide an intuitive understanding of how GPs work. Most readers of this review will, however, be more familiar with least-squares regression. In this section we will show that GPR can be thought of as a generalisation of least-squares regression, allowing for correlated noise (or signals) in the data. Conversely, least-squares regression, as traditionally presented, is a special case of GPR, where the covariance matrix is assumed to be purely diagonal, and the variances associated with each observation are known \textit{a priori}.

\subsubsection{Revisiting least squares}\label{sec:lsq}

Once again, we consider $N$ observations of a variable $\mathbf{y} = \{ y_i\}_{i=1, \ldots, N}$, taken at times $\mathbf{t} = \{ t_i\}$, with associated measurement uncertainties $\boldsymbol{\sigma}=\{\sigma_i\}$. We wish to compare these to a model function $m(t,\meanparams)$ controlled by parameters $\meanparams=\{\theta_j\}_{j=1,\ldots,M}$. In most astrophysical applications, we are interested in estimating some or all of those parameters. In least-squares regression, we minimize the quantity
\begin{equation}
  \chi^2 \equiv\sum_{i=1}^N (y_i-m_i)^2/\sigma^2_i,
\end{equation}
where $m_i \equiv m(t_i,\meanparams)$, with respect to $\meanparams$. Where does this come from?

Let us assume that the observations are given by
\begin{equation}
  y_i=m(t_i,\meanparams)+\epsilon_i,
\end{equation}
where $\epsilon_i$ is the measurement error, or noise, on the $i^{\rm th}$ observation. Furthermore, let us assume that $\epsilon_i$ is drawn from a Gaussian distribution with mean $0$ and variance $\sigma_i^2$:
\begin{equation}
  p(\epsilon_i)=\mathcal{N}(0,\sigma_i^2) \equiv \frac{1}{\sqrt{2\pi} \sigma_i} \exp\left(-\frac{\epsilon_i^2}{2\sigma_i^2} \right),
\end{equation}
then the \emph{likelihood} for the $i^{\rm th}$ observation is simply
\begin{equation}
  \mathcal{L}_i (\meanparams) \equiv p(y_i|\meanparams)=\mathcal{N}(m_i,\sigma_i^2) =\frac{1}{\sqrt{2\pi} \sigma_i} \exp\left[-\frac{(y_i-m_i)^2}{2\sigma_i^2} \right].
\end{equation}
We also assume that the noise is uncorrelated, or white, meaning that the $\epsilon_i$'s are drawn independently of each other from their respective distributions. Then, the likelihood for the whole dataset $\mathbf{y}$ is merely the product of the likelihoods for the individual observations:
\begin{equation}
  \mathcal{L} (\meanparams) \equiv p(\mathbf{y}|\meanparams) = \prod_{i=1}^N \mathcal{L}_i = \prod_{i=1}^N \left\{ \frac{1}{\sqrt{2\pi} \sigma_i}
  \exp\left[-\frac{(y_i-m_i)^2}{2\sigma_i^2} \right] \right\}.
\end{equation}
From this one can readily see that $\ln \mathcal{L} = \mathrm{constant} - 0.5 \chi^2$, where the constant depends only on the $\sigma$'s. Thus, if the $\sigma$'s are known, maximizing $\mathcal{L}$ is equivalent to minimizing $\chi^2$. In other words, least-squares regression yields the Maximum Likelihood Estimate (MLE) of the parameters under the assumption of white, Gaussian noise with known variance.

\subsubsection{Link to Gaussian Process Regression}\label{sec:gp-lsq-link}
Let us now re-write the likelihood in matrix form:
\begin{equation}
\label{eq:gp_like}
  \mathcal{L}(\meanparams,\hyperparams) = 
  \frac{1}{\sqrt{2\,\pi\,|\mathbf{K}|}} \exp \left(-\tfrac{1}{2}(\mathbf{y}-\mathbf{m})^{\mathrm{T}} \mathbf{K}^{-1}
  (\mathbf{y}-\mathbf{m}) \right),
\end{equation}
where, as before, the mean vector $\mathbf{m}$ has elements $m_i=m(t_i,\meanparams)$ and $\mathbf{K}$ is a purely diagonal ($N$, $N$) matrix with elements $K_{ij} = \delta_{ij} \sigma_i^2$ ($\delta_{ij}$ being the discrete Kronecker delta function). This is, of course, the \emph{covariance matrix} of the model. The notation $|\mathbf{A}|$ refers to the determinant of the matrix $\mathbf{A}$, and $\mathbf{A}^{-1}$ to its matrix inverse.

Now, instead of assuming that the covariance matrix takes this very specific form, let us allow a more flexible covariance model:
\begin{equation}
  K_{ij} = k(t_i,t_j,\hyperparams) + \delta_{ij} \sigma_i^2,
\end{equation}
where $k$ is a \emph{covariance function}, or \emph{kernel function}, controlled by parameters $\hyperparams$. The result is a GP, and its likelihood is still given by \autoref{eq:gp_like}. Depending on the choice of kernel function and parameters, the covariance matrix can now have non-zero off-diagonal elements, allowing us to explicitly model correlated noise or stochastic signals in the data.
Note that the likelihood depends not only on the argument of the exponential, but also on the determinant of the covariance matrix, which therefore needs to be evaluated explicitly. This term acts as a built-in Occam's razor, automatically penalising more complex models.

The kernel function $k$ encodes our beliefs about the stochastic, or random, element of the model, in just the same way as the mean function $m$ encodes our beliefs about the deterministic component of the model. For example, in many circumstances, we would expect that two observations taken close together in time should be more strongly correlated than observations taken further apart, and we would use a decreasing function of the time interval $\dt=|t_i-t_j|$ to represent that. One of the most widely used covariance functions is the squared exponential
\begin{equation}
\label{eq:kse}
  k(\dt;\,\hyperparams) = \amplitude^2\,\exp\left(-\frac{\dt^2}{2\,\lengthscale^2}\right).
\end{equation}
This gives rise to smooth (infinitely differentiable) random functions with variance $\amplitude^2$ and characteristic length-scale $\lambda$. We present other commonly used kernel functions and discuss how to select one in \autoref{sec:kernels}.

\subsection{Inference with a GP}\label{sec:gp-inf}

\begin{figure}[ht]
  \centering
  \includegraphics[width=\linewidth]{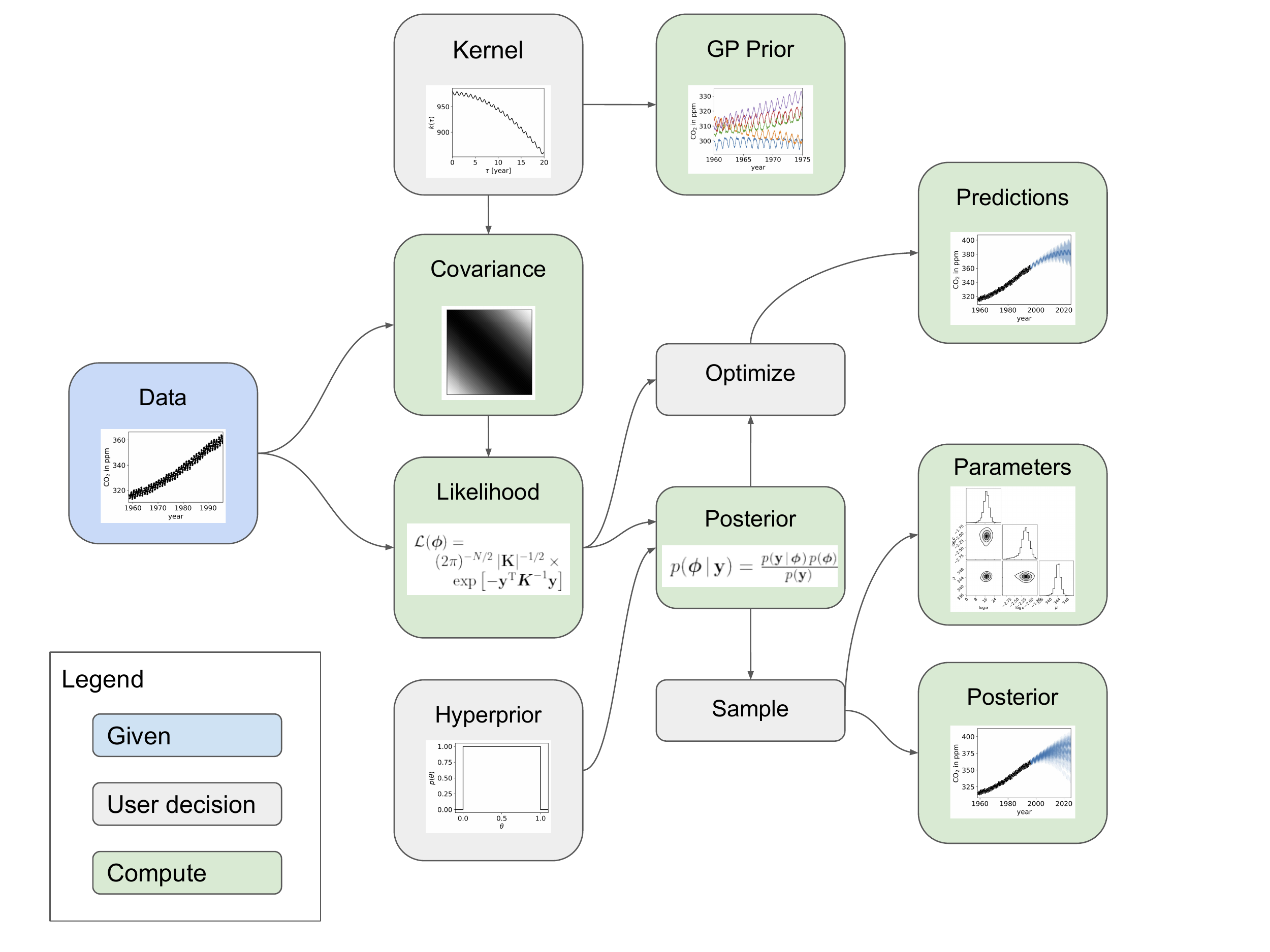}
  \caption{Schematic representation of a typical GPR workflow. Given a dataset (blue box) and some modelling choices (grey boxes, \autoref{sec:choices}), the mathematical framework presented in \autoref{sec:basics} can be used (green boxes) to evaluate the likelihood and posterior distribution over the hyperparameters which can be optimised or sampled. Note that, for simplicity, we have assumed a zero mean function in this figure. We can also condition the GP on the data to predict observations at locations where we do not yet have observations. Several of the plots in this figure are taken from the 
  tutorial on hyperparameter optimization included in the \project{george} package documentation (\url{https://george.readthedocs.io}), and show the composite GP model proposed by \citet{gpml} for the Mauna Loa CO$_2$ dataset of \citet{MaunaLoaCO2}.}  
  \label{fig:GPR_workflow}
\end{figure}

Now that we know how to evaluate the likelihood (\autoref{eq:gp_like}), we are ready to perform \textit{inference}, that is to use observations to update our prior beliefs about the system we are observing. The overall Bayesian inference workflow for GPR regression is illustrated schematically in \autoref{fig:GPR_workflow}. Given a dataset (blue, left panel), our first task is typically to select a kernel function, along with an educated initial guess for its hyperparameters ($2^{\rm nd}$ column, top panel). These choices are discussed in more detail in \autoref{sec:choices}, and should be guided by a combination of domain knowledge (what we know about the process that generated the data), practical considerations (e.g.\ ease of implementation and computational cost) and diagnostics based on the data themselves (e.g.\ examining the auto-correlation function and/or the power spectral density of the data). Given a kernel function and hyperparameters, we can draw samples from the GP prior ($3^{\rm rd}$ column, top panel), which can serve as sanity check for the choice of kernel function (in the sense that, measurement errors aside, samples from the GP prior should ideally look qualitatively similar to the observations). 

We are now in a position to compute the GP covariance matrix for the data ($2^{\rm nd}$ column, $2^{\rm nd}$ panel), and to evaluate the likelihood of the model ($2^{\rm nd}$ column, $3^{\rm rd}$ panel), which we can then optimise with respect to the hyperparameters ($3^{\rm rd}$ column, $2^{\rm nd}$ panel) using standard optimization methods; this is called \emph{training} the GP. Often, however, we would incorporate priors on the hyperparameters ($2^{\rm nd}$ column, bottom panel), which we select once again based on domain knowledge and practical considerations. We can then evaluate the \emph{posterior} distribution ($3^{\rm rd}$ column, $3^{\rm rd}$ panel):
\begin{equation}
p(\meanparams,\hyperparams|\mathbf{y})=\frac{p(\mathbf{y}|\meanparams,\hyperparams)p(\meanparams,\hyperparams)}{p(\mathbf{y})},
\end{equation}
where the denominator $p(\mathbf{y})=\int p(\mathbf{y}|\meanparams,\hyperparams)p(\meanparams,\hyperparams) \mathrm{d}\meanparams \mathrm{d}\hyperparams$ is the marginal likelihood, or model evidence, and is often omitted from calculations since it is independent of the hyperparameters, but becomes relevant for model comparison. Once again, we can optimise this posterior with respect to the hyperparameters ($3^{\rm rd}$ column, $2^{\rm nd}$ panel). For a given set of hyperparameters (for example, the set which maximises either the likelihood or the posterior), the GP prior can be conditioned on the observations to make predictions ($4^{\rm th}$ column, top panel); this is called \emph{conditioning} the GP, and is discussed in \autoref{sec:pred}. 

However, in astronomical applications we often need to estimate the full posterior distribution over the hyperparameters. 
Generally we cannot do this analytically, so we resort to sampling methods such as Markov Chain Monte Carlo (MCMC) or nested sampling ($3^{\rm rd}$ column, bottom panel). The samples can be used to estimate posterior distributions for individual hyperparameters of interest ($4^{\rm th}$ column, $2^{\rm nd}$ panel). For example, we might be using a GP to take into account correlated noise when fitting an otherwise deterministic model to some data, but we are not interested in the noise \emph{per se}, only on the impact it has on the parameters of the mean function, $\meanparams$. We would then \emph{marginalise} (integrate over, or project) the posterior samples over the \emph{nuisance parameters} $\hyperparams$. Some samplers also enable us to compute the model evidence, which can be used to compare different kernel functions (see \autoref{sec:model_selection} for a discussion). Finally,  posterior samples can also be used in a Monte Carlo fashion to compute predictions that incorporate the uncertainties on the hyperparameters ($4^{\rm th}$ column, bottom panel).

\subsection{Making predictions}
\label{sec:pred}

An important use case for GPR is as a statistically principled interpolation method. We wish to ``learn'' an unknown function that gave rise to some data, in order to make predictions for some new set of inputs. 
Importantly, a GP model provides not a point estimate but a full probability distribution for the function at any desired location(s) in the input domain. This allows for robust uncertainty propagation (though there are some important caveats we will touch upon in \autoref{sec:caution}), and can also motivate strategies for active sampling (deciding when or where to make new observations).

\subsubsection{The predictive equations}\label{sec:predeq}

Given some existing observations $\mathbf{y}$, taken at times $\mathbf{t}$, how do we make predictions at some new set of times $\mathbf{t}_\star$? We are after $p(\mathbf{y}_\star|\mathbf{y})$, the \textbf{conditional} probability distribution for $\mathbf{y}_\star$ given $\mathbf{y}$. In GPR this is also known as the \emph{predictive distribution}, because it is often used to extrapolate a time-series dataset forwards. The ``magic'' of GPR is that the predictive distribution is also Gaussian, and its mean and covariance are given by simple analytic relations:
\begin{equation}
  \label{eq:pred}
  \mathbf{f}_\star = \mathbf{m}_\star + \mathbf{K}_\star^{\mathrm{T}} \, \mathbf{K}^{-1} \, (\mathbf{y} - \mathbf{m}) ~\mathrm{~~~and~~~}
  \mathbf{C}_\star =   \mathbf{K}_{\star\star} - \mathbf{K}_\star^{\mathrm{T}} \, \mathbf{K}^{-1} \, \mathbf{K}_\star,
\end{equation}
where we have introduced the vector $\mathbf{m}_\star \equiv m(\mathbf{t}_\star;\meanparams)$ and the matrices $\mathbf{K}_\star$ and $\mathbf{K}_{\star\star}$, with elements
\begin{equation}
  \label{eq:Kstar}
    [\mathbf{K}_\star]_{ij} = k(t_i,\mathbf{t}_{\star,j}; \hyperparams) ~~~~~ \text{and} ~~~~~
    [\mathbf{K}_{\star\star}]_{ij} = k(\mathbf{t}_{\star,i},\mathbf{t}_{\star,j};\hyperparams).
\end{equation}
Note that, as real observations are always noisy, $\mathbf{K}$ generally includes a white noise term ($\delta_{ij} \sigma^2_i$), but $\mathbf{K}_\star$ does not. The white noise term is usually omitted from (the diagonal elements of) $\mathbf{K}_{\star\star}$, in which case the predictive covariance $\mathbf{C}_\star$ accounts only for the uncertainties in the function inferred with the GP model. When making predictions that represent hypothetical additional observations taken with the same setup as the training set, however, the white noise term should be included in $\mathbf{K}_{\star\star}$. 

\subsubsection{Properties of the predictive distribution}
A closer look at \autoref{eq:pred} reveals some important properties.
First, a GP is a \textit{linear predictor}\footnote{%
  There is some subtlety with respect to the non-linear mean function so, in this section, we set the mean to zero or, equivalently, consider the \emph{residuals} away from the mean model as the ``data''.}.
The predictive mean for a specific time $t_\star$ can be written as a linear combination of the observations:
$f_\star = \mathbf{w}^{\mathrm{T}} \, \mathbf{y}$, where $\mathbf{w} = \mathbf{K}^{-1} \, \mathbf{k}_\star$ and $\mathbf{k}_\star=k(\mathbf{t},t_\star,\hyperparams)$. It can also be written as
a linear combination of covariance functions centred on the training points:
$f_\star = \mathbf{\alpha}^\mathrm{T} \,  \mathbf{k}_\star$ where $\mathbf{\alpha} = \mathbf{K}^{-1} \, \mathbf{y}$.
These linearity properties are very important to understand the behaviour of GPs, as they shed light on the relationship between GPR and other types of models, including standard linear models with very large numbers of free parameters \citep[see e.g.][for a more detailed discussion]{2021PASP..133i3001H}.

Second, the predictive covariance is independent of the data. $\mathbf{C}_\star$ depends only on the \emph{locations} $\mathbf{t}$ of the observations, not on their values $\mathbf{y}$. This has important consequences for observation planning: if we know the covariance function $k$ and its parameters, we can decide when to observations to optimise our predictions at a given (set of) time(s). However, in practice we rarely know the hyperparameters \textit{a priori}, and the predictive posterior distribution marginalised over the hyperparameters \textit{does} depend on the observations.

Finally, as $\mathbf{K}$ is positive semi-definite, so is $\mathbf{K}^{-1}$. Therefore, $\mathbf{k}_\star^{\mathrm{T}} \, \mathbf{K}^{-1} \, \mathbf{k}_\star \ge 0,$ and $\mathrm{Var}(\mathbf{f}_\star) \leq k(t_\star,t_\star)$. This is as we would expect: obtaining more data should only ever improve the accuracy of our predictions.

\subsubsection{Cautionary notes}
\label{sec:caution}
It is important to note that the behaviour of $\mathbf{f}_\star$ is \emph{not} the same as that of individual samples from the predictive distribution. Typically,  $\mathbf{f}_\star$ tends to be smoother than individual samples. This should be borne in mind when displaying GPR results or using them in subsequent analysis.

It is also important to note that the predictive variance accounts for the imperfect ability of the specific model under consideration to explain the data, but not for the choice of model (i.e. the choice of mean and kernel functions and their parameters). We discuss how to choose a kernel function and fit for its parameters in the next section.

\section{GAUSSIAN PROCESS MODELLING DECISIONS}
\label{sec:choices}

In the previous section, we presented an overview of GP methods, and the key mathematical details. In this section, we will dive deeper into some of the practical decisions that arise when using GPs.
The two core elements of a GP model are the mean function $m(t;\,\meanparams)$, and the ``kernel'' or ``covariance'' function $k(t_i,\,t_j;\,\hyperparams)$. In contrast to the way GP models are typically presented in the machine learning literature, in astrophysics they will often---but not always---include non-trivial mean functions.
For example, in the case study discussed in \autoref{sec:transit}, the mean function $m(t;\,\meanparams)$ is a physical transit model that is a function of the orbital parameters of the system, and includes a realistic limb-darkening model \citep{2002ApJ...580L.171M}.
However, in this review we won't discuss the mean function in detail, focusing instead primarily on the kernel function, since that is unique to GP modelling.
All this being said, in our experience, new users of GP models will often focus and worry more than necessary about the choice of kernel function for their problem.
As with any probabilistic modelling problem, there are several well-defined workflows for motivating, selecting, and validating your choice of kernel function.
In this section we walk through this process in detail.

\subsection{Gaussian Process Covariance Functions}
\label{sec:kernels}

\begin{textbox}[htb]
  \section{Kernel function}
  The ``kernel'' or ``covariance'' function, typeset here as $k(t_i,\,t_j;\,\hyperparams)$ is a parametric description of the covariance between two data points $t_i$ and $t_j$. This function takes a pair of inputs (here $t_i$ and $t_j$) as arguments, and is controlled by a set of parameters $\hyperparams$, typically referred to as ``hyperparameters''.

  \vspace{1.5em}

  \section{Hyperparameters}
  In a GP model, the kernel function is typically parameterized by a set of parameters that we will label as $\hyperparams$ and refer to as ``hyperparameters''.
  Formally, the ``parameters'' of a GP are the (unknown) ``true'' values of the process at the observed times, but the magic of Gaussians means that these parameters can be marginalised over in closed form, leaving the hyperparameters as the parameters of interest.
\end{textbox}

The kernel function can be any positive scalar function that gives rise to a positive semi-definite covariance matrix over the input domain, but some will be more useful than others.
Given this large decision space, you may be wondering how to choose the right kernel function for your specific problem.
If you are very lucky, you may be able to motivate your model using physics.
This can be approached from two directions.
On one hand, some commonly used kernel functions have a specific physical interpretation---for example, the solution to a stochastic ordinary differential equation---that would naturally motivate their use in certain circumstances.
On the other hand, if your physical model is specified by its \emph{power spectrum}, this can be recast as a GP model with a specific kernel function.

Even if you do not have a formal physics-based justification for your model, you may be able to identify the key scales of your problem and design a kernel function that captures these features.
The usual approach to this problem is to take sums and products of commonly used kernel functions to construct a set of models that have the desired covariance structure, and then combine or select between these choices.
For example, if the problem of interest is expected to involve only a single non-periodic timescale, you could list all the two-parameter non-periodic kernel functions and use a numerical model selection technique as described below.

\subsubsection{Standard kernel functions}\label{sec:standard-kernels}

Some popular kernel functions are listed in \autoref{tab:kernels}, and some other choices are discussed in Chapter 4 of \citet{gpml}. We already met the 
squared exponential kernel function, defined in \autoref{eq:kse}. Its  hyperparameters are $\hyperparams = \{\amplitude,\,\lengthscale\}$, where $\amplitude$ controls the amplitude (output scale) and $\lengthscale$ the length scale (input scale) of the functions. Due to its simplicity, this is the most widely-used kernel for modelling smooth functions of unknown shape.

Another example that is commonly used in the astrophysics literature \citep{2012MNRAS.419.3147A,2014MNRAS.443.2517H} is the following ``quasi-periodic'' kernel function
\begin{equation}
\label{eq:kqp}
  k(\dt;\,\hyperparams) = \amplitude^2\,\exp\left(-\frac{\dt^2}{2\,{\lengthscale_1}^2} -\Gamma\,\sin^2\left[\frac{\pi\,\dt}{\lengthscale_2}\right] \right)
\end{equation}
with $\hyperparams = \{\amplitude,\,\lengthscale_1,\,\lengthscale_2,\,\Gamma\}$, which has a period of $\lengthscale_2$, and a decoherence timescale of $\lengthscale_1$. The hyperparameter $\Gamma$ controls the extent to which the periodic component of the signal resembles a simple sinusoid or is more complex.

It is worth noting that, in any practical application, these kernel functions are not generally used on their own.
Instead, more expressive models are designed by combining these models as discussed in \autoref{sec:kernel-ops}.

\begin{figure}[ht]
  \centering
  \script{samples.ipynb}
  \begin{minipage}[t]{0.3\linewidth}
    \includegraphics[width=\linewidth]{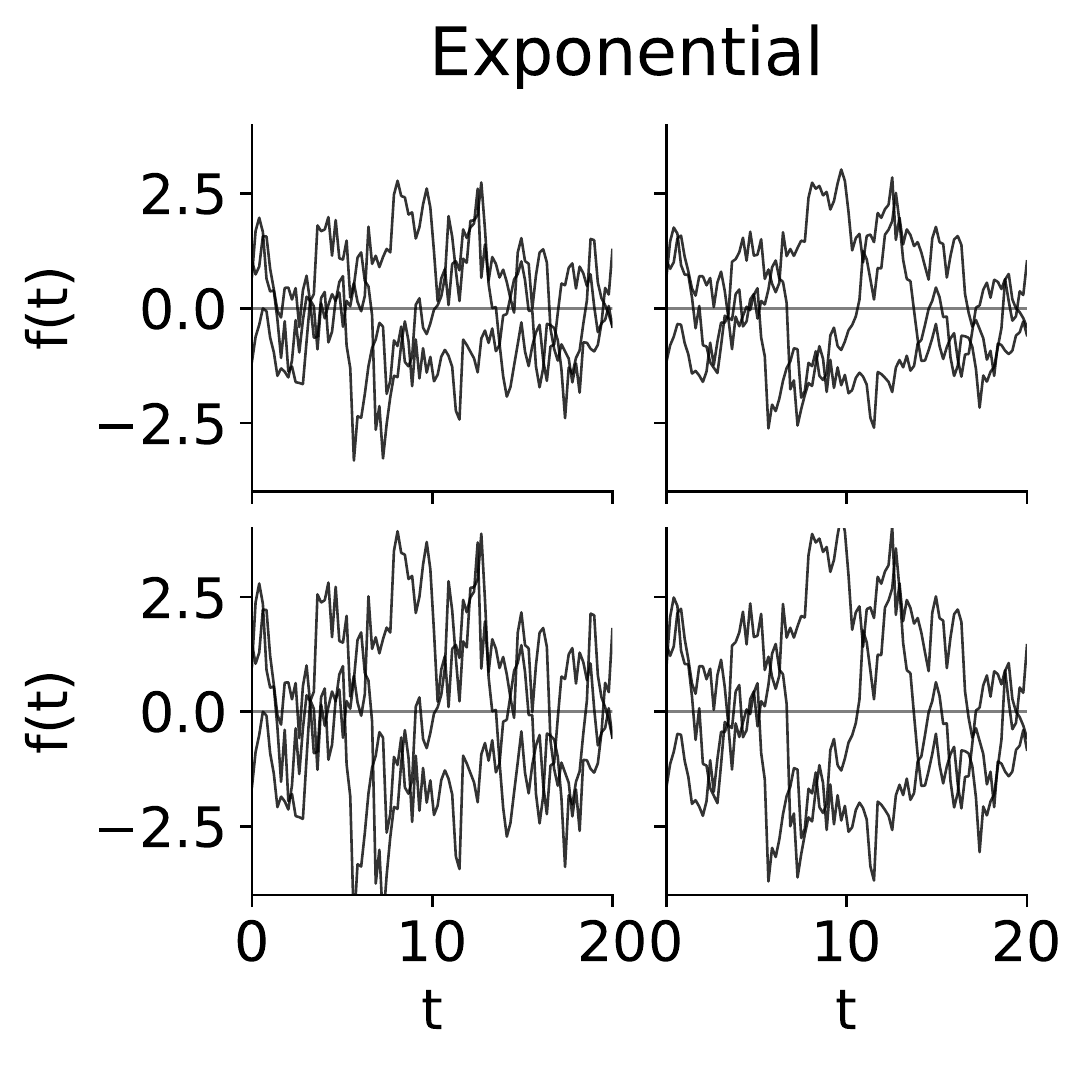}
  \end{minipage} \hfill
  \begin{minipage}[t]{0.3\linewidth}
    \includegraphics[width=\linewidth]{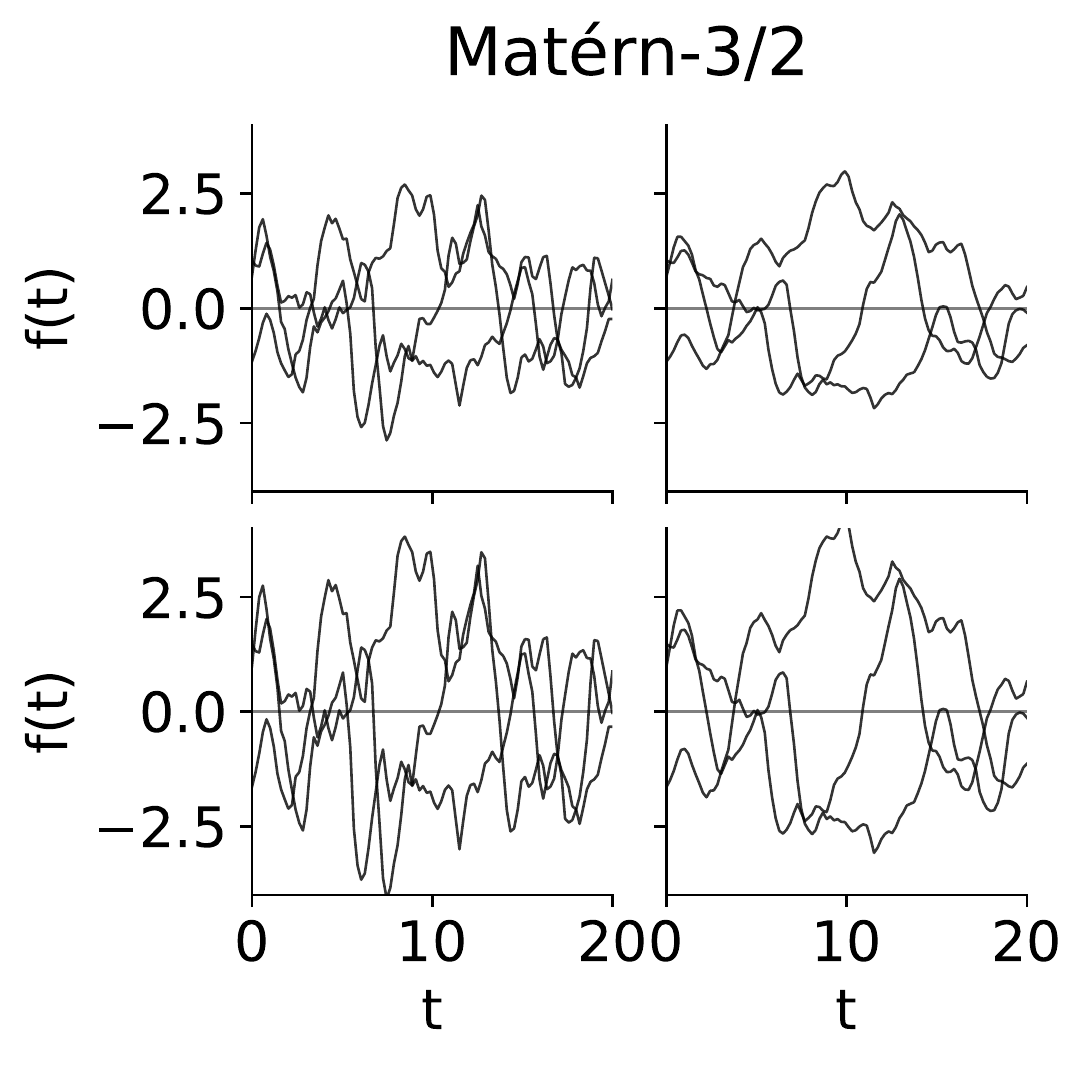}
  \end{minipage} \hfill
  \begin{minipage}[t]{0.3\linewidth}
    \includegraphics[width=\linewidth]{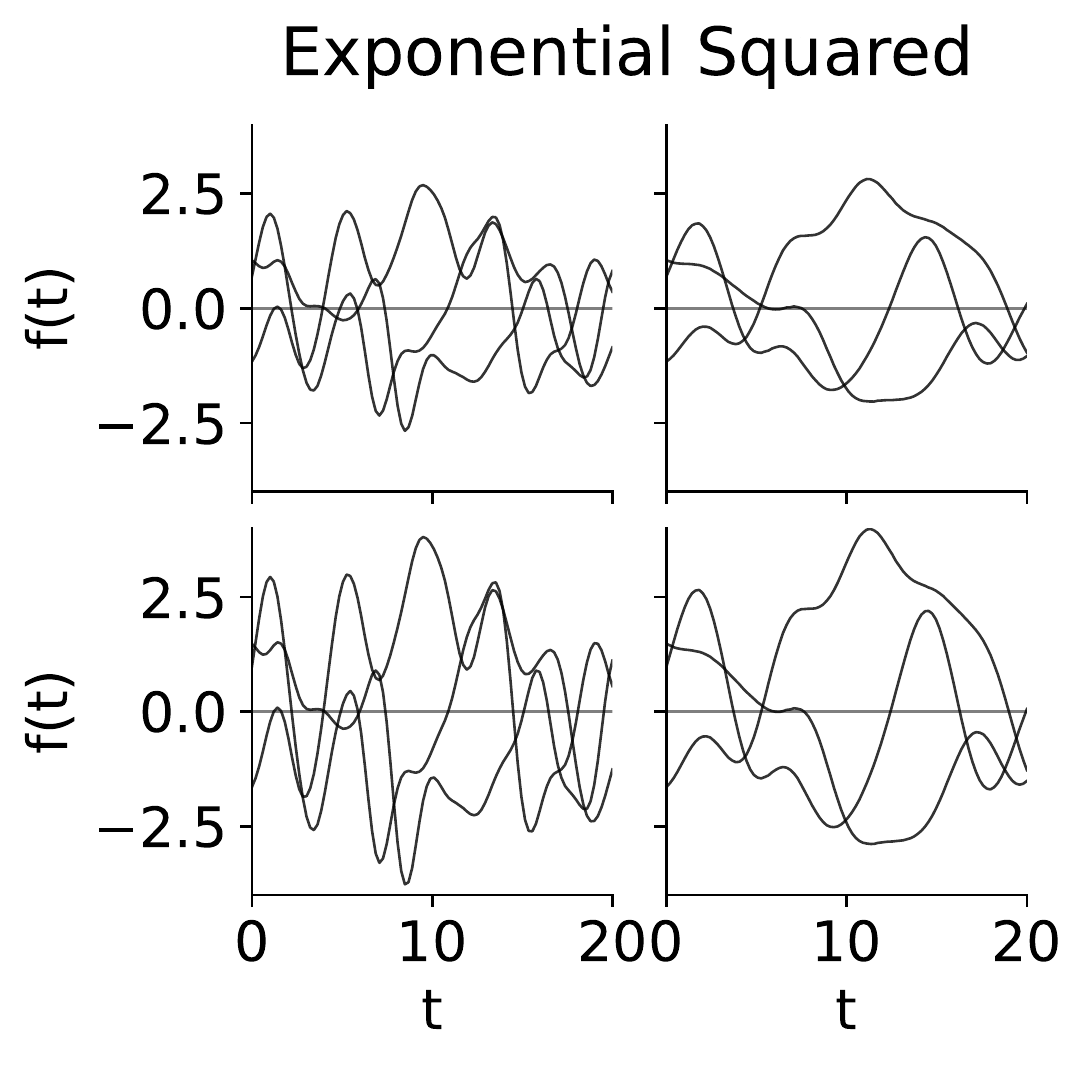}
  \end{minipage}
  \caption{Prior samples for three different classes of kernel functions: (a) exponential, (b) Mat\'ern-3/2, and (c) exponential squared.
  In each sub-figure, the length scale $\lambda$ of the kernel increases from left to right, and the amplitude $\alpha$ increases from top to bottom.
  One thing to notice in this figure is that these kernel functions differ in their smoothness properties.
  Specifically, the exponential kernel is not mean-square differentiable, while the exponential squared kernel is infinitely differentiable.
  This can be seen qualitatively in this figure.}
  \label{fig:samples}
\end{figure}

When selecting a kernel function, it can be useful to generate samples from this implied prior distribution over functions to get a qualitative sense of the properties of the kernel.
In practice, this is done by choosing a grid of times $\boldsymbol{t} = \{t_i\}$, evaluating the elements of the covariance matrix
\begin{equation}
  K_{ij} = k(t_i,\,t_j;\,\hyperparams) \quad,
\end{equation}
and then generating a multivariate Gaussian sample with this covariance.
As an example, \autoref{fig:samples} shows several prior samples for three different kernel functions from \autoref{tab:kernels}, with a range of length scales $\lengthscale$ and amplitudes $\amplitude$.
In this figure, we can see some qualitative differences between the kernel functions---namely that the samples become ``smoother'' from left to right---and we can see how the characteristic input and output scales of the allowed functions change with the hyperparameters.

A key feature of a kernel function that defines its qualitative behaviour is its mean square (MS) differentiability.
We will not formally define MS differentiability---interested readers are instead referred to Section~4.1.1 of \citet{gpml} for a more detailed discussion---but we want to highlight that MS differentiability is important because it defines the ``smoothness'' of the functions that can be modelled by the GP.
It can be seen in \autoref{fig:samples} that the exponential squared kernel function, which is infinitely MS differentiable, is significantly ``smoother'' than the exponential kernel function, which is not MS differentiable.

\begin{armarginnote}[]
  \entry{MS differentiability}{mean square differentiability}
\end{armarginnote}

\begin{table}[ht]
  \caption{Some kernel functions commonly used in the astrophysics literature. }
  \label{tab:kernels}
  \begin{center}
    \begin{tabular}{@{}l|l@{}}
      \hline
      Name                           &  Representation$^{\rm a}$                                    \\
      \hline
      Constant                       & $\amplitude^2$                                                     \\
      Squared Exponential$^{\rm b}$  & $e^{-(\dt/\lengthscale)^2/2}$                                              \\
      Exponential$^{\rm c}$          & $e^{-\dt/\lengthscale}$                                                  \\
      Mat\'ern-3/2                   & $\left(1 + \sqrt{3}\dt/\lengthscale\right)e^{-\sqrt{3}\dt/\lengthscale}$             \\
      Mat\'ern-5/2                   & $\left(1 + \sqrt{5}\dt/\lengthscale +5(\dt/\lengthscale)^2/3\right)e^{-\sqrt{5}\dt/\lengthscale}$  \\
      Rational quadratic             & $\left( 1 + \frac{\dt^2}{2 \gamma\lengthscale^2} \right)^{-\gamma}$ \\
      Cosine                         & $\cos 2\pi\dt/\lengthscale$                                                  \\
      Sine Squared Exponential       & $\exp\left(-\Gamma\sin^2\pi\dt/\lengthscale\right)$                      \\
      Stochastic Harmonic Oscillator$^{\rm d}$ & $\cos\left(\sqrt{1-\beta^2}\frac{\dt}{\lengthscale}\right) + \frac{\beta}{\sqrt{1-\beta^2}}\sin\left(\sqrt{1-\beta^2}\frac{\dt}{\lengthscale}\right)$                          \\
      \hline
    \end{tabular}
  \end{center}
  \begin{tabnote}
    $^{\rm a}$in each case, $\dt$ is defined as $\dt = \left|t_i - t_j\right|$, and Greek letters indicate hyperparameters;
    $^{\rm b}$``radial basis function'';
    $^{\rm c}$``Ornstein--Uhlenbeck'', ``damped random walk'' or ``Mat\'ern-1/2'';
    $^{\rm d}$stochastically-driven damped simple harmonic oscillator \citep{celerite}.
  \end{tabnote}
\end{table}

\subsubsection{Operations on kernel functions}\label{sec:kernel-ops}

As discussed above, more flexible kernel functions are often constructed using products and sums of the standard kernels listed in \autoref{tab:kernels}.
Besides, addition and multiplication, other operations can be used to impose structure on the standard kernels.
For example, linear operations like scalar multiplication, more general affine transformations, differentiation, or integration can all be used to develop new kernel functions.

For example, there are good arguments \citep{2012MNRAS.419.3147A} to expect that the radial velocity time series of a spotted rotating star is related to the photometric time series via its time derivative.
More formally, assume we have two time series---one with observations of some function $f(t)$ and another with observations of $\dot{f}(t)$, where the dot indicates the derivative with respect to time---and the function $f(t)$ is modelled as a GP with some covariance function $k(t_i,t_j;\hyperparams)$. The covariance between an observation of $f(t)$ and one of $\dot{f}(t)$ is
\begin{equation}
  \mathrm{Cov}\left[f(t_i),\dot{f}(t_j)\right] = \left.\frac{\partial k(t_i,t;\hyperparams)}{\partial t}\right|_{t=t_j},
\end{equation}
and the covariance between two observations of $\dot{f}(t)$ is
\begin{equation}
  \label{eq:cross-deriv-covar}
  \mathrm{Cov}\left[\dot{f}(t_i),\dot{f}(t_j)\right] = \left.\frac{\partial^2 k(t,t^\prime;\hyperparams)}{\partial t\,\partial t^\prime}\right|_{t=t_i,t^\prime=t_j} \quad.
\end{equation}
This means that we can compute the GP model for observations of both sets of observations and their covariances.
One technical point which is worth noting here: for \autoref{eq:cross-deriv-covar} to define a valid covariance function, the base kernel $k(t,t^\prime;\hyperparams)$ must be mean square differentiable, as mentioned briefly in the previous section.
Therefore, the exponential kernel, for example, should not be used for modelling derivative observations since it does not have the appropriate properties.

\begin{figure}[ht]
  \centering
  \script{kernel-ops.ipynb}
  \includegraphics[width=0.7\linewidth]{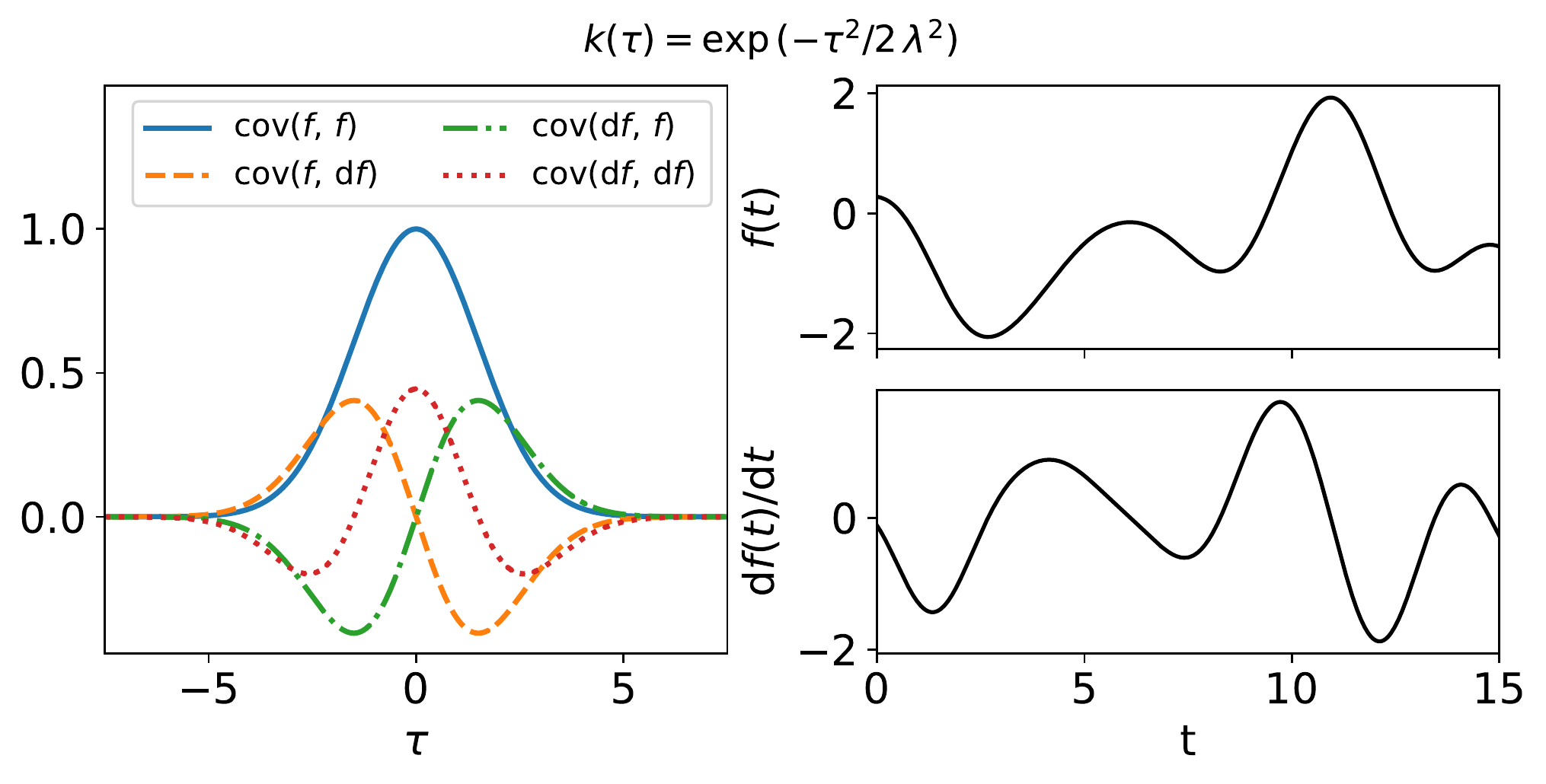}
  \includegraphics[width=0.7\linewidth]{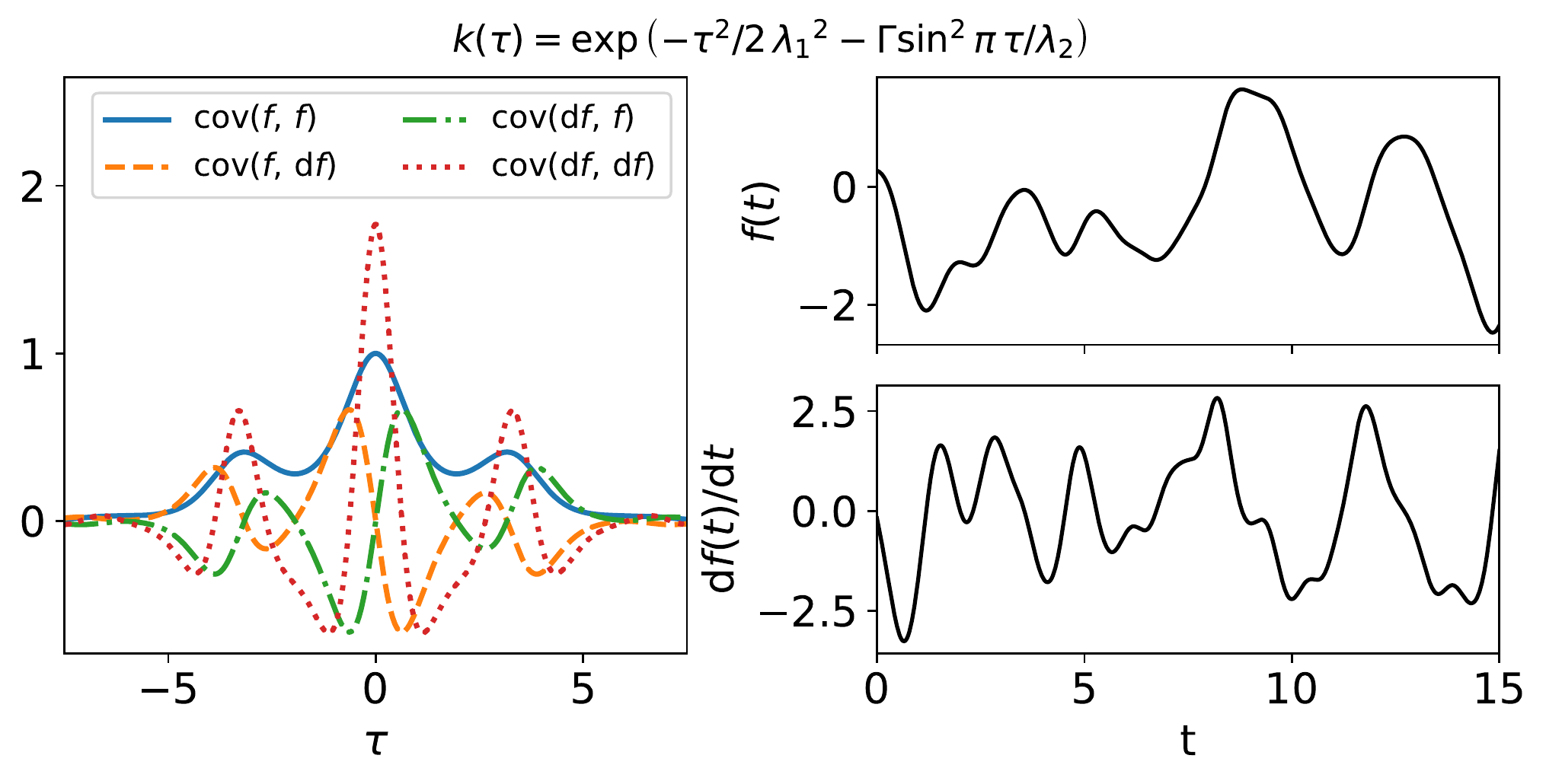}
  \caption{A demonstration of the effects of differentiation on a GP model.
  Each row shows the same figures a different base kernel: \emph{(top)} a squared exponential kernel ($\lengthscale = 1$), and \emph{(bottom)} the product of a squared exponential kernel ($\lengthscale_1 = 2.5$) with a sine squared exponential kernel ($\Gamma = 1$; $\lengthscale_2 = 3.5$).
  In each row, the leftmost panel shows the kernel function and its first and second time derivatives, which are used to evaluate the covariance matrix for a GP model for a process and its derivative.
  The right-hand panel shows a realization from the process ($f(t)$, top) and its time derivative ($\dot{f}(t)$, bottom) generated by jointly sampling from the GP including the covariance between $f(t)$ and $\dot{f}(t)$.}
  \label{fig:kernel-ops}
\end{figure}

\autoref{fig:kernel-ops} illustrates the elements of GP models with different kernel functions for observations of a time series and its time derivative.
The left panels show the covariance functions for all pairwise permutations of $f(t)$ and $\dot{f}(t)$, and the right panels show samples of the function and its derivative.
The use of this type of GP kernel is discussed in more detail in \autoref{sec:RVact}, with applications to radial velocity time series observations of exoplanets.

Integration or convolution also have similar properties.
For example, real time series observations are made with finite exposure time $\Delta$, and therefore, if the underlying stochastic process $f(t)$ is modelled as a GP with covariance $k(t_i,t_j;\hyperparams)$, the observations are actually of
\begin{equation}
  f_\mathrm{int}(t) = \int_{t-\Delta/2}^{t+\Delta/2} f(t^\prime)\,\mathrm{d}t^\prime \quad,
\end{equation}
which has a covariance of
\begin{equation}
  k_\mathrm{int}(t_i,t_j;\hyperparams) = \int_{t_i-\Delta/2}^{t_i+\Delta/2}\int_{t_j-\Delta/2}^{t_j+\Delta/2} k(t,t^\prime;\hyperparams)\,\mathrm{d}t\,\mathrm{d}t^\prime \quad,
\end{equation}
which can be evaluated in closed form for some standard kernel functions.
Another context where integrated GPs have recently been used for astrophysics (albeit not for time series), is as a model of the Galactic dust distribution \citep{2022arXiv220206797M}.

\subsubsection{Multivariate GPs}

Since this review is focused on GPs for \emph{time domain} astronomy, we primarily consider univariate GPs with just \emph{time} as the input coordinate.
However, we would be remiss to neglect discussing multivariate GPs altogether.
Most of our discussion so far applies to multivariate datasets without any change, but there are some subtleties that are worth mentioning here.

When working with univariate GPs, it is relatively straightforward and unambiguous to compute the ``distance'' between two points $t_i$, and $t_j$: $\rho(t_i,t_j) = \dt = \left|t_i - t_j\right|$.
For multivariate inputs $\boldsymbol{x}$, more care must be taken to define a sensible distance metric $\rho(\boldsymbol{x}_i,\boldsymbol{x_j})$.
Common choices include the $d$-th norm $\rho_d(\boldsymbol{x}_i,\boldsymbol{x_j}) = \left|\left|\boldsymbol{x}_i - \boldsymbol{x}_j\right|\right|_d$ for $d=1$ or $d=2$, but other metrics---for example, the Haversine distance when the elements of $\boldsymbol{x}$ are coordinates on a sphere---might be better motivated for a particular use case.

When the input coordinates have some geometric interpretation (e.g., spatial coordinates) simple isotropic distance metrics, like those presented so far, are generally well-justified.
However, when the input coordinates are heterogeneous with, for example, different units, it is often more appropriate to use a more general distance metric.
In particular, you may want to fit for a different length scale parameter $\lengthscale_d$ for each input dimension $d$:
\begin{equation}\label{eq:ard-distance}
  \rho(\boldsymbol{x}_i,\boldsymbol{x}_j) = \sqrt{\sum_{d=1}^D \left(\frac{x_{id} - x_{jd}}{\lengthscale_d}\right)^2}
\end{equation}
or even include support for covariance between different input dimensions:
\begin{equation}
  \rho(\boldsymbol{x}_i,\boldsymbol{x}_j) = \sqrt{\boldsymbol{x}_i\,\boldsymbol{M}^{-1}\,\boldsymbol{x}_j}
\end{equation}
where $\boldsymbol{M}$ is a $D\times D$ positive definite matrix of hyperparameters.

Kernels with a different length scale parameter for each dimension (\autoref{eq:ard-distance}) are commonly referred to as ``automatic relevance determination'' \citep[ARD;][]{gpml} kernels, the idea being that the data can inform the importance of each input dimension when describing the observations.
In practice, such a kernel is often used in conjunction with a ``shrinkage'' prior on the hyperparameters, favouring large values of $\lengthscale_d$ as discussed further in \autoref{sec:transit_fit}.
\begin{armarginnote}[]
  \entry{ARD}{Automatic Relevance Determination}
\end{armarginnote}

A common class of multivariate GP model for time domain astronomy are datasets with multiple parallel covariant time series.
For example, multi-band time series produced by surveys like PanSTARRS or LSST, or radial velocity time series with parallel activity indicators.
One useful way to specify these datasets are to define the inputs as $\boldsymbol{x_i} \equiv (t_i,\ell_i)$ where $t_i$ the time of the $i$-th observation and $\ell_i$ is a ``label'' for which time series the $i$-th observation is drawn from.
In this case, one common choice of kernel function is:
\begin{equation}\label{eq:multiband}
  k(\boldsymbol{x}_i,\boldsymbol{x}_j;\hyperparams) = \boldsymbol{a}_{\ell_i}^\mathrm{T} \boldsymbol{a}_{\ell_j} k_0(t_i,t_j;\hyperparams) \quad,
\end{equation}
where $k_0(t_i,t_j;\hyperparams)$ is a standard one-dimensional kernel, and the set of $\{\boldsymbol{a}_d\}_{d=1}^D$ are also hyperparameters of the model.
This class of kernels includes the models commonly used for interpolating multi-band photometric observations of transients \citep[e.g.,][]{2020ApJ...905...94V, 2021ApJS..255...24V}.
In the case where the data are rectangular (i.e., all input dimensions are observed at all times), the covariance matrix defined by \autoref{eq:multiband} will have Kronecker structure \citep[e.g.,][]{2015arXiv151101870W, 2020AJ....160..240G}, which can be exploited to improve the computational performance of the GP model.

\subsubsection{Physically motivated kernels}
\label{sec:physical}
Physical processes that are expected to be GPs appear frequently in astrophysics.
For example, a physical model that is specified by a parametric form for its ``power spectrum'' in the Fourier domain (think, the cosmic microwave background, or solar-like asteroseismic oscillations), could \emph{equivalently} be specified by its covariance function (or kernel function) in real space.
It is common practice in astrophysics to analyse datasets with these models by first Fourier transforming the data, and then fitting these transformed data using a physical model for the power spectrum.
There is a rich literature on this topic, discussing the motivation for this approach, and the rigorous propagation of measurement uncertainty for these datasets.
These analyses could also be performed using GPR without first transforming the data, but this has typically been avoided since na\"ive implementation of GPR would be significantly more computationally expensive, or even completely intractable.

There are, however, many cases where analysis in the time domain is preferred---at least in principle---over a Fourier analysis.
This is especially true when working with short, significantly unevenly sampled, or otherwise heterogeneous datasets, or when combined with non-stationary functions that cannot be compactly represented in the frequency domain. 

Another class of GPs that are commonly used for astrophysics are stochastic differential equations \citep[see][for a detailed discussion of these models]{Sarkka:2019}, although this terminology is not standard in the literature.
The most widely-used model in this class is the Ornstein--Uhlenbeck \citep{1930PhRv...36..823U}, or ``damped random walk'' process \citep[e.g.,][]{2009ApJ...698..895K, 2010ApJ...708..927K, 2010ApJ...721.1014M, 2012A&A...546A..89B, 2020AJ....160..265H}.
These models---or the generalized ``continuous-time autoregressive moving average'' (CARMA) process \citep[e.g.,][]{2014ApJ...788...33K, celerite, 2022ApJ...936..132Y}---are commonly used as a physically-motivated model for the stochastic time-variability of active galactic nuclei (AGN).
Although this fact is not always noted in the literature, these models are examples of GPs with specific kernel function choices.
Besides their physical interpretability, these models can also be evaluated with computationally efficient and scalable algorithms \citep[e.g.,][]{2014ApJ...788...33K, celerite}, which can be crucial for practical applications, as we discuss in \autoref{sec:fast}.

Asteroseismology is another domain where Fourier space models are common, and recently there has been some work using time-domain GPR to analyse these datasets \citep[e.g.,][]{2017AJ....154..254G, 2018ApJ...865L..20F}.
The kernel functions used for these analyses are typically parameterized by the frequency and amplitude of the oscillation modes, but the likelihood is evaluated using \autoref{eq:gp_like} in the time domain.
Time-domain GP models for asteroseismic oscillations can be particularly useful when combined with a mean model that is compactly specified in the time domain \citep[e.g., a transiting planet,][]{2017AJ....154..254G}, or when the time sampling of the data is strongly non-uniform with complicated windowing effects \citep{2018ApJ...865L..20F}.
Like the AGN models discussed above, these asteroseismic GP models can typically be evaluated efficiently using scalable algorithms.
So far, the applications of GPR to asteroseismology have been generally restricted to simple effective models with a single wide mode used to capture the power spectrum of the oscillations, but it is conceivable that more complex models could be applied to fully capture the structure of the power spectrum.

While the models discussed so far in this section have strong physical motivations, it is also common to apply GPR using effective kernel models with physically-motived \emph{parameters}.
For example, GPs have been used to measure the rotation periods of stars in photometric time series using a quasi-periodic kernel function, where one parameter of the model can be interpreted as the stellar rotation period \citep{2018MNRAS.474.2094A}.
These models perform well and can be used to formally propagate the observational uncertainty to the inferred rotation period.
Recently, some more physically interpretable kernel functions have been developed for measuring stellar rotation, and mapping stellar surfaces using photometric time series \citep{2021AJ....162..124L}, but these models have so far seen limited use because of their computational cost.


\subsection{Hyperparameter Inference}\label{sec:inference}

A key component of all the covariance functions discussed above is that they are all \emph{parameterized} by a set of hyperparameters.
In most cases, you won't have \emph{a priori} knowledge for how the values of these hyperparameters should be set.
Instead, their values will need to be numerically tuned or incorporated into a larger inference scheme.

In the astrophysics literature, the most common approach for taking this uncertainty into account---and the method that we advocate for here---is to treat the hyperparameters directly as parameters of the model.
In other words, instead of just fitting for the parameters of the mean model $\meanparams$, we can simultaneously fit for both $\meanparams$ and the hyperparameters $\hyperparams$.
In \autoref{sec:gp-lsq-link}, we defined the likelihood for a GP model and, in \autoref{sec:gp-inf}, we sketched the procedure used for Bayesian inference with such a model.
The likelihood function defined in \autoref{eq:gp_like} is a function of both $\meanparams$ and $\hyperparams$, and we can use that function as an objective for a non-linear optimization routine to find the maximum likelihood parameter values, or in a Markov chain Monte Carlo \citep[MCMC; e.g.][]{2018ApJS..236...11H} procedure to marginalise over the hyperparameters, and propagate their uncertainty to constraints on the parameters of the mean model.

An important point here is that, since many data analysis procedures in astrophysics include a step like the ones listed above, the use of a GP likelihood does not significantly change the processing.
In fact, we like to say that the GP likelihood can be used as a drop-in replacement for anywhere you are currently using a ``chi-squared'' objective.
There are some practical reasons why things are not necessarily this simple (for example, computational cost, as described below), but the sentiment stands.

In this review, we won't go into too many details about the inference algorithms, but throughout the text, we will regularly use the \textsf{BFGS} gradient-based, non-linear optimization routine \citep{Nocedal:1999, scipy} to find the maximum likelihood parameter values
\begin{equation}
  \meanparams_\star,\hyperparams_\star = \argmax_{\meanparams,\hyperparams} \,\mathcal{L}(\meanparams,\hyperparams) = \argmax_{\meanparams,\hyperparams} \,\log p(\ydata\,|\,\meanparams,\hyperparams) \quad.
\end{equation}
Another common inference technique used in astrophysics---and in this review!---is the use of MCMC or nested sampling \citep{skilling2006nested} to generate posterior samples
\begin{equation}
  \meanparams,\hyperparams \sim p(\meanparams,\hyperparams\,|\,\ydata)
\end{equation}
that can be used to marginalise over some subset of the parameters, and estimate the uncertainty on the parameter values.
All the examples in this review are implemented using the \textsf{NumPyro} probabilistic inference library, and all the MCMC examples use the gradient-based No U-Turn Sampler (NUTS) algorithm \citep{Hoffman:2014}.

\autoref{sec:quasar} and \autoref{sec:transit}, and specifically \autoref{fig:quasar} and \autoref{fig:transit}, show the results of worked examples of an MCMC-based GPR workflow, without and with (respectively) a nontrival mean function.

\subsection{Model assessment, validation, and selection}
\label{sec:model_selection}

Given the wide array of possible kernel functions described in \autoref{sec:kernels}, it can be important to assess the performance of your model and the relevant choices.
This includes both assessing the choice to use a GP in the first place, and the specific choice of kernel function.
It is important to note that there is nothing fundamentally different about GPs in this context when compared to other models for data, so that means that many methods that you may already use for model selection and validation in other contexts will also apply when using GPs.
That being said, within the astrophysics literature formal probabilistic model checking has had limited use, and GPs do come with some specific technical complications, therefore we will discuss some examples of model validation and selection techniques that have been used for GPs.

When it comes to selecting between different possible kernel functions, the approach that you take may depend on your specific research goals.
For example, in many cases, including the transiting exoplanet example in \autoref{sec:transit}, the main parameters of physical interest may be the parameters of the mean model, and the GP is simply an effective model for stochastic nuisances.
In this case, it may be sufficient to demonstrate that the inferred results are not significantly inconsistent for different choices of kernel function.

Other common use cases, like the time delay example in \autoref{sec:quasar}, primarily require good predictive performance for the GP model.
In these cases, methods like cross-validation \citep[e.g.,][]{gelfand1992model} or posterior predictive assessment \citep[e.g.,][]{gelman1996posterior} can be used to evaluate different choices of kernel function.

\begin{figure}[ht]
  \centering
  \script{assessment.ipynb}
  \includegraphics[width=\linewidth]{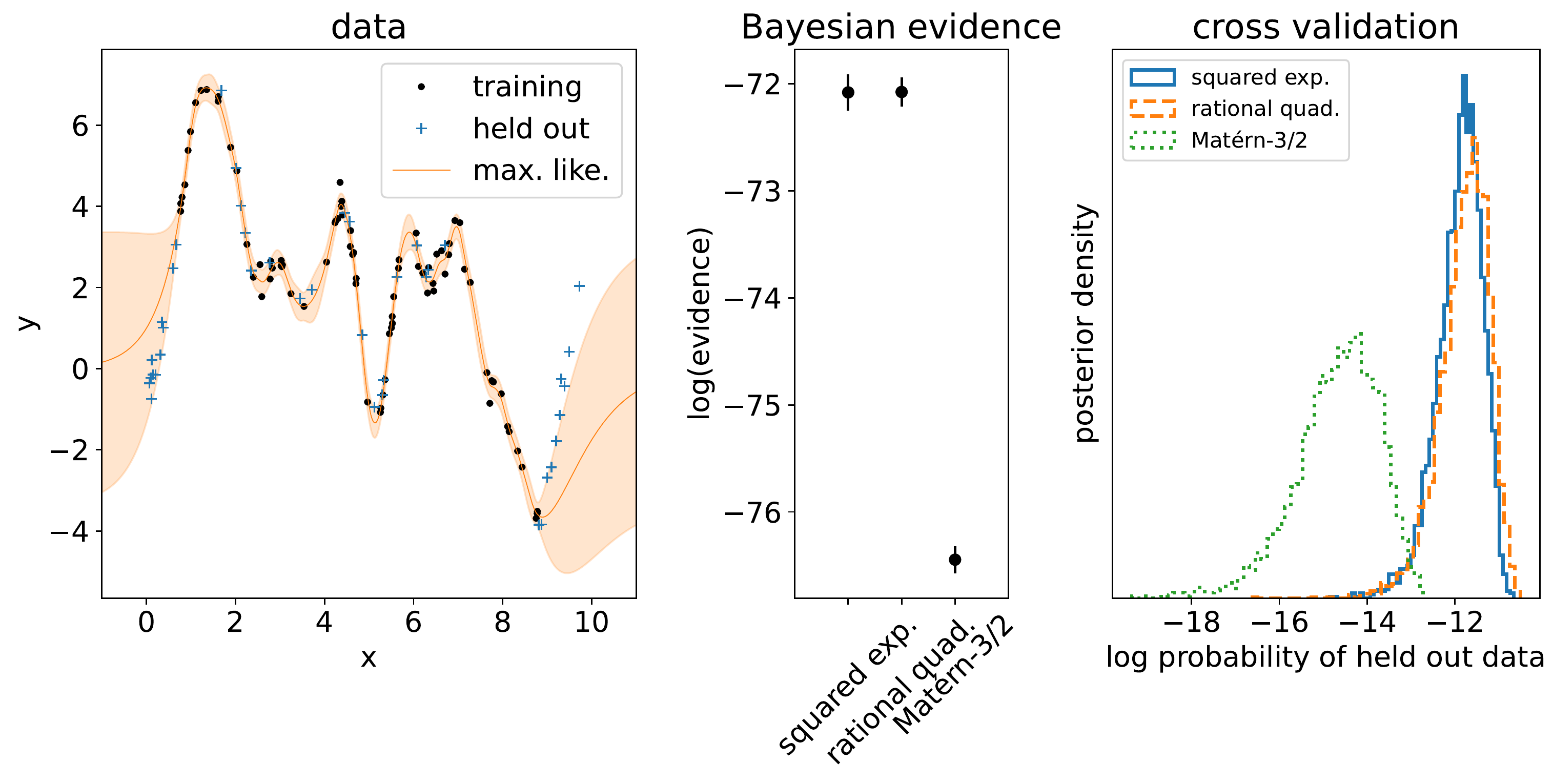}
  \caption{A demonstration of model assessment using data simulated from a GP model using a squared exponential kernel with known parameters.
  \emph{(left)} The data simulated data are shown as black dots, with some held out for cross validation indicated by blue crosses.
  The maximum likelihood estimate of the GP's predictive distribution based only on the black points is indicated with the orange $1-\sigma$ contours.
  \emph{(middle)} The value of the Bayesian evidence integral numerically estimated using nested sampling and all the data in the left panel (including the data held out for cross-validation), is shown for each kernel function.
  \emph{(right)} The posterior distribution of the held out log probability is plotted for each kernel function.}
  \label{fig:assessment}
\end{figure}

To demonstrate these approaches, \autoref{fig:assessment} shows the results of performing a model comparison between three different kernel functions applied to a simulated dataset.
The simulated data were generated from a GP model with a squared exponential kernel with known parameters, and we aim to compare the performance of three kernels: (1) the squared exponential kernel, (2) the Mat\'ern-3/2 kernel, and (3) the rational quadratic kernel.
The simulated dataset is shown in the left panel of \autoref{fig:assessment}.

First, using the full dataset, we compute the Bayesian evidence integral for each of these model choices using a nested sampling algorithm implemented in the \textsf{jaxns} package \citep{jaxns}.
The Bayesian evidence integral is defined as:
\begin{equation}
  Z(H) \equiv p(\ydata\,|\,H) = \int p(\meanparams,\hyperparams\,|\,H)\,p(\ydata\,|\,\meanparams,\hyperparams,H)\,\mathrm{d}\meanparams\,\mathrm{d}\hyperparams
\end{equation}
where $H$ indicates the modelling choices, in this case the choice of kernel function.
The Bayesian evidence is frequently used in the astrophysics literature as an ingredient in model selection procedures \citep[see][for a more detailed discussion]{2008ConPh..49...71T}.
Nested sampling is an algorithm for numerically estimating this integral, and we apply it to produce estimates of $Z$ for each choice of kernel $H$, and plot these results in the middle panel of \autoref{fig:assessment}.
In this figure, it is clear that the squared exponential and rational quadratic kernels are indistinguishable under this metric, while the Mat\'ern-3/2 kernel is somewhat disfavoured. 

Another popular method for model selection that is less commonly used in the astrophysics literature is cross validation.
Cross validation is designed to assess the predictive performance of the model, and it proceeds by holding out some data, fitting the rest of the data, and then computing the likelihood of the held out data conditioned on the fit results.
These steps can then be repeated for different held out samples.
For a GP model, the likelihood of the held out data can be computed using the predictive distribution discussed in \autoref{sec:pred}.
In particular, the likelihood is the following multivariate Gaussian
\begin{equation}\label{eq:held-out-like}
  p(\ydata_\star\,|\,\ydata,\meanparams, \hyperparams) = \mathcal{N}(\boldsymbol{f}_\star,\boldsymbol{C}_\star)
\end{equation}
where $\boldsymbol{f}_\star$ and $\boldsymbol{C}_\star$ are defined in \autoref{eq:pred}, noting (importantly!) that the (squared) observational uncertainties on the held out data should be included on the diagonal of $\boldsymbol{C}_\star$.

In this example, we use MCMC to fit the data shown as black dots in the left panel of \autoref{fig:assessment}, holding out the data points indicated by blue crosses.
At each step in the MCMC, we evaluate the likelihood in \autoref{eq:held-out-like} for the held out data conditioned on the training data and the model parameters.
In the rightmost panel of \autoref{fig:assessment}, we plot the posterior distribution of the held out log probability to show that, like with the evidence integral, the squared exponential and rational quadratic kernels are indistinguishable, while the Mat\'ern-3/2 kernel is disfavoured.
In this case we hold out a single fixed subset of the data, but it is common practice to repeat this process for several unique held out subsets, and average these results.
There is some freedom in the choice of held out data, and it is not trivial to select the best procedure for correlated time series data.
In this example, we held out some contiguous data points at the beginning and end of the time series to assess the model's performance when extrapolating, and then held out a random subset of the remaining data points to validate the bulk statistics of the process.
For GP models, simple ``leave-one-out'' or other uniform sampling schemes are unlikely to be sufficient for assessing the correlation structure of the data, and care should be taken to select validation datasets that capture the details of interest.

Besides the model comparison tests discussed so far, it can also be useful to assess the quality of fit in a more absolute sense.
Like with the other assessment tools, there is not anything particularly unique about GP models when it comes to model checking, although it is generally non-trivial to quantify goodness of fit metrics in all but the simplest cases \citep[e.g.,][]{gelman1995bayesian}.
Our preferred approach to model checking is to use graphical posterior predictive checks \citep[see][for example]{bayes-workflow}.
The idea behind posterior predictive checks is that the observed data $\ydata$ should not be an extreme outlier with respect to synthetic data $\ydata_\star$ sampled from the posterior:
\begin{equation}
  \ydata_\star \sim p(\ydata_\star\,|\,\ydata) = \int p(\ydata_\star\,|\,\meanparams,\hyperparams)\,p(\meanparams,\hyperparams\,|\,\ydata)\,\mathrm{d}\meanparams\,\mathrm{d}\hyperparams \quad.
\end{equation}
Operationally, a single sample of $\ydata_\star$ is generated by drawing a sample of $\meanparams$ and $\hyperparams$ from the posterior, and then drawing $\ydata_\star$ from the sampling distribution $p(\ydata_\star\,|\,\meanparams,\hyperparams)$ conditioned on the sampled values of $\meanparams$ and $\hyperparams$.
Datasets in astrophysics are typically sufficiently high dimensional that it is not feasible to visualize the full distribution of $\ydata_\star$, and we instead plot scalar statistics of the data, such as the mean, variance, or skew.
If the model is appropriate for the problem at hand, the same statistic computed on the observed data should be consistent with the posterior predictive distribution.

\section{GAUSSIAN PROCESSES IN TIME-DOMAIN ASTRONOMY}
\label{sec:uses}

In this section we discuss a range of applications of GPR to time-domain datasets in the astronomical literature. The present-day popularity of GPs in astronomy precludes any attempt at an exhaustive review. Instead, we have selected examples which showcase the power and flexibility of the method. No special meaning should be ascribed to the order in which these applications are discussed, beyond the fact that we started with the application domains we are most familiar with.

\subsection{Stars and exoplanets}
\label{sec:stars_planets}

It is no coincidence that the recent uptake of GPR in astronomy coincides with the meteoritic rise of exoplanet studies. Almost all exoplanet detections and related discoveries are made indirectly, and involve detecting small and/or short-lived signals in time-domain datasets: transits in light curves, Keplerian signals in Radial Velocity (RV) or astrometry time-series. These are invariably affected by, and often buried in, correlated noise or nuisance signals, making GPR a natural choice.


\subsubsection{Transit searches}
\label{sec:transit_det}

Following the discovery of the first transiting exoplanet \citep{2000ApJ...529L..41H,2000ApJ...529L..45C}, numerous ground-based photometric monitoring surveys were set up (or re-purposed) to search for planetary transits. Early estimates of their yield \citep[see e.g.][]{2003ASPC..294..361H} ran into the hundreds of planets per year, but these soon turned out to be highly optimistic. While surveys such as the Optical Gravititational Lensing Experiment (OGLE) did indeed discover numerous transit-like events
\citep[see e.g.][]{2002AcA....52....1U}, only a handful were ultimately confirmed as bona-fide transiting planets, and the process took years. 
The vast majority of early transit candidates were later diagnosed as diluted or grazing stellar eclipses based on radial velocity follow-up.
Intriguingly, a small fraction turned out to be spurious detections, despite having relatively high signal-to-noise ratio (SNR). Indeed, \citet{2006MNRAS.373..231P} noted that all the confirmed OGLE planets had very high SNRs, and went on to show that the SNRs for all candidates had been over-estimated due to the presence of correlated, or red, noise in the light curves. They proposed an empirical method to evaluate the red noise on transit time-scales from the individual light curves (analogous to the method proposed by \citealt{prh92a} to evaluate the correlation matrix of a quasar light curve), and then a prescription for adjusting the detection threshold accordingly. In effect, they were searching for a way to model the covariance of the data. This is precisely what GPR allows, but the methodology was not known in the exoplanet community at the time.

GP models have been used on occasion to \textit{de-trend} light curves from transit surveys prior to running transit searches, i.e.\ to remove variability on timescales significantly longer than a transit \citep[see e.g.][]{2016ApJS..226....7C}, and are used frequently after detection to model the out-of-transit baseline alongside the transit signal itself (see \autoref{sec:transit_fit}). In principle, a GP could also be used to model out-of-transit variations (intrinsic or instrumental) alongside transits as part of a detection pipeline. Doing this simultaneously rather than sequentially allows more flexible models to be used for the out-of-transit variations, and avoids the risk of corrupting the transit signal by using an excessively aggressive filter at the detection stage, as demonstrated by \citet{2015ApJ...806..215F} for \textit{K2} data. In that case, the primary source of out-of-transit variability is instrumental systematics (see \autoref{sec:transit_fit}), which are to some degree common to all light curves in a given observing run. \citet{2015ApJ...806..215F} model these systematics as a linear combination of the first 150 principal components of the ensemble of light curves, alongside a simple box-shaped transit model, varying the latter's period, phase and duration.

In other space-based transit surveys such as \textit{CoRoT}, \emph{Kepler} and \emph{TESS}, the systematics are less prominent, so the dominant source of out-of-transit variations is intrinsic stellar variability, which occurs mainly on timescales longer than transits, and can therefore be filtered out quite effectively without removing the transits. However, in some cases, for example for rapidly rotating and magnetically active stars, the separation between transits and stellar signal becomes less clean, significantly reducing the performance of sequential de-trending approaches (irrespective of the algorithm used, \citealt{wotan}).
In such cases, a simultaneous modelling approach could improve detection performance significantly. Since the nuisance signal is specific to each star, it cannot be modelled using other light curves, but a GP is a credible alternative. Such an approach could also offer a small but significant enhancement in the sensitivity of future missions such as \textit{PLATO} to transits of habitable planets around Sun-like stars (which are very shallow and relatively long). The main issue would be computational cost: transit surveys monitor $10^4$--$10^5$ stars at a time, with $10^3$--$10^5$ observations per run, and the transit search must be run over a fine grid of periods and phases. Doing this with standard GPR methods would be impractical. However, now fast and scalable GP solvers are available, this becomes a more feasible proposition.

\subsubsection{Transit and eclipse modelling}
\label{sec:transit_fit}

In practice, GPR has been much more widely applied to transit modelling than to detection \textit{per se}. The detailed depth, timing, duration and shape of a transit depends on the planet-to-star radius ratio $R_{\rm p}/R_\star$, the system scale $a/R_\star$ (where $a$ is the orbital semi-major axis), the impact parameter $b$, the orbital period $P$ and the time of transit centre, $T_0$ as well as the limb-darkening profile of the star. High-precision observations of one or more transits of a given planet can be used to infer these parameters, which has a wide range of scientific applications. The most obvious is to estimate the planet size (and hence, given a mass estimate, its bulk density and composition). The system scale is directly related to the stellar density \citep{2003ApJ...585.1038S}, which can be used for sanity checks to ensure the transits are indeed of planetary origin.
Small departures from strictly periodic timing, known as Transit Timing Variations (TTVs), probe dynamical interactions between multiple planets in a given system, and can be used measure the planets' masses and eccentricities and to reveal the presence of additional, non-transiting planets \citep{2005Sci...307.1288H}. Wavelength-dependent measurements of the transit depth probe the effect of the planet's atmosphere on the starlight that filters through it, and hence allow us to access the atmospheric composition \citep{2000ApJ...537..916S}. The depth of any secondary eclipse (when the planet passes behind the star) depends on the planet-to-star flux ratio, and can therefore be used to measure an emission or reflection spectrum, while its timing yields strong constraints on the orbital eccentricity. All of this requires precise and accurate measurements of the transit parameters, for which any correlated noise must be accounted for explicitly, as we demonstrated using a simulated example in \autoref{sec:transit}.

To address this, \citet{2009ApJ...704...51C} proposed a wavelet-based method to model correlated noise in transit light curves, which can be seen as a special case of GPR with a covariance function belonging to the exponential family, but formulated in terms of wavelets. Rather than modelling the covariance in the time domain, the Power Spectral Density (PSD) of the noise is assumed to be of the form $P(f) \propto f^{-\gamma}$.
Noting that the wavelet transform of such a process gives rise to a nearly diagonal covariance matrix, \citet{2009ApJ...704...51C}  derived an expression for the likelihood that can be computed in $\mathcal{O}(N)$ operations. 
This method has been used widely since its publication to model transit observations from space missions which stare continuously at a given field, such as \textit{Kepler} or \textit{TESS}.
\begin{armarginnote}
  \entry{Colours of noise}{White, pink and red noise correspond to $\gamma=0$, $1$ and $2$, respectively.}
\end{armarginnote}
However, the use of a wavelet transform requires regular time sampling, which precludes its application to datasets with irregular sampling or significant data gaps, for example from ground-based, space telescopes in low-Earth orbit such as Hubble.

This limitation is significant, as most datasets for transit and eclipse spectroscopy are unevenly sampled with significant data gaps. Transit spectroscopy involves measuring minute changes in the depth of the primary transit or secondary eclipse as a function of wavelength, either using successive photometric observations through different filters or using a spectrograph (see \citealt {2018haex.bookE.100K} for a review). The signal of interest is the wavelength dependence of the transit depth, or the depth of the eclipse, which are of order $10^{-4}$ and $10^{-3}$ respectively, in  the most favourable cases. Even when observed from space, these signals are typically dwarfed by instrumental systematics. The pointing jitter and thermal relaxation of the telescope as it orbits the Earth causes the target star to move on the detector, typically by a fraction of a pixel. As the sensitivity of the detector varies from pixel to pixel (the ``flat-field'') and between the centre and edges of each pixel, this motion causes spurious variations in the recorded flux from the target. Although we expect the measured flux to depend on ``housekeeping" variables such as the satellite orbital phase, telescope attitude, the centroid of the image or the locus and angle of the spectrum of the detector, or the temperatures of various parts of the instrument, a physically motivated model for the form of this dependence is generally lacking. \textit{Ad hoc} parametric (e.g.\ polynomial) models are problematic, as the choice of model inputs and functional form is arbitrary, yet can drastically alter the resulting exoplanet spectrum \citep{2011MNRAS.411.2199G}.

To address this, \citet{2012MNRAS.419.2683G} proposed a GPR framework to model the systematics in space-based transit observations. The aforementioned housekeeping variables are treated as multidimensional inputs to a squared exponential GP whose mean is the transit signal, allowing one to marginalise over broad families of systematic models without assuming a specific functional form for them, all the while propagating the resulting uncertainties on the physical parameter of interest, namely the (wavelength-dependent) planet-to-star radius ratio. The kernel used was a squared exponential kernel with an automatic relevance determination (ARD) distance metric (\autoref{eq:ard-distance}), this allows one to try including a wide range of housekeeping data into the fit; in principle only those which are genuinely relevant will have an effect on the result.
GP-based systematics models of this kind were later extended to eclipse spectroscopy, where the parameter of interest is the planet-to-star flux ratio, which controls the eclipse depth. Among other successes, this approach has enabled the first measurement of the wavelength-dependent albedo of a hot Jupiter \citep{2013ApJ...772L..16E}, helped resolve early controversies surrounding the treatment of systematics in Spitzer observations \citep{2015MNRAS.451..680E} and
led to the first unambiguous detection of a thermal inversion in an exoplanet emission spectrum \citep{2017Natur.548...58E}. Using simulated data, \citet{2014MNRAS.445.3401G} showed that GPs outperform parametric models when the true form of the systematics is unknown, but also that the most robust results overall are obtained by marginalising over families of both parametric and GP models.

Despite these theoretical advantages and practical successes, the use of GPR to analyse low-resolution transit and eclipse spectra remains confined to a relatively small subset of the corresponding community. Parametric models (in some cases marginalising over families thereof, \citealt{2016ApJ...819...10W}) remain the most widely-used approach for Hubble Space Telescope observations, and Pixel-Level-Decorrelation (PLD) for Spitzer observations \citep{2015ApJ...805..132D}. Possible explanations for this include computational cost, as well as the fact that these other methods are perceived as easier to implement and interpret. Another likely reason, at least in the case of HST transmission spectra, is that GP systematics models tend to produce exoplanet spectra with larger uncertainties than parametric ones. This does not mean that they are less accurate, however, merely that they make a less restrictive set of assumptions about the functional form of the systematics. As discussed by \citet{2014MNRAS.445.3401G} and demonstrated in the appendix of \citet{2018AJ....156..283E}, when the true functional form of the systematics is unknown, choosing the ``wrong'' systematic model can lead to overconfident estimates of the transit depth, which is more problematic for the robust interpretation of the results than the enlarged uncertainties produced by a more agnostic GP model.

One general shortcoming of current GP-based systematics models with multidimensional inputs is that the uncertainties on the input variables are ignored. This could be remedied by treating the housekeeping variables as noisy observations of latent GP variables, and modelling them alongside the observed fluxes, as done by \citet{2016MNRAS.462..726A} in the context of photometric redshift determination, but we are not aware of published attempts to do this in practice to date. \begin{armarginnote}
  \entry{Latent variable}{A random variable which is never actually observed.}
\end{armarginnote}
In spectroscopic observations, whether using GPs or parametric models, it has become standard practice to model the ``white" light curve first (obtained by integrating the spectrum over the full wavelength range at each time-step). ``Coloured'' light curves, extracted in individual wavelength bins, are then divided by the best-fit systematics model derived from the white light curve, before being modelled further. To our knowledge, no attempt has been made to model the wavelength dependence of the systematics directly.

On the other hand, GPR has become quite widely used for modelling correlated noise in single-band transit observations. One striking example is the case of transiting planet candidates discovered by \textit{Kepler} around giant stars. By comparing the stellar densities derived from transit modelling to those expected from independent estimates, \citet{2014ApJ...788..148S} argued that many of these candidates, including the confirmed planet Kepler-91, might be false positives. However, red giant light curves contain stochastic variability on timescales of hours due to granulation. In the case of Kepler-91, \citet{2015ApJ...800...46B} showed that the apparent density discrepancy disappears when this granulation signal is modelled using a GP.

\subsubsection{Systematics removal in the presence of stellar variability}
GP-based models have also been used to correct instrumental systematics in data from the \textit{K2} space mission, which used the \textit{Kepler} satellite to perform an Ecliptic plane survey after the failure of two of its reaction wheels. During the \textit{K2} observations, the satellite underwent significant roll-angle variations, causing the stars to move on the detector by more than a pixel over a timescale of hours. The satellite thrusters were fired every $\sim 6$\,h to return the spacecraft to its nominal attitude, but the resulting drift caused significant changes in the measured stellar fluxes, due to the detector inter- and intra-pixel variations, and to changes in the contamination of the photometric aperture by neighbouring targets as well as in aperture losses. These can be modelled effectively using a GP with a squared exponential covariance function depending on the roll angle \citep{2015MNRAS.447.2880A, 2016ApJS..226....7C} or the star's 2-D position on the detector \citep{2016MNRAS.459.2408A}. In these approaches, a second, time-dependent term is added to the GP covariance function to represent the target star's intrinsic variability. This not only improves the fit but allows the position-dependent systematics to be evaluated separately and thus removed while preserving intrinsic variability. The use of a time-dependent GP term improves the photometric performance over other widely-used methods for correcting \textit{K2} systematics \citep[e.g.,][]{2014PASP..126..948V} when the stellar variability is significant and/or occurs on timescale similar to the roll angle variations. Ultimately, the best overall photometric precision for \textit{K2} was achieved by combining PLD to model the position-dependent systematics with a time-dependent GP to model intrinsic variability \citep{everest1,everest2}.

\begin{textbox}[ht]
\section{Composite GPs}
A composite (additive or multiplicative) GP is a powerful way to separate different components of a time-series dataset, whether these components depend on different input variables (as in the instrumental systematics examples discussed in this section) or not. To evaluate the predictive distribution for a particular component of a composite GP, one simply evaluates $\mathbf{K}_\star$ and $\mathbf{K}_{\star\star}$ in \autoref{eq:pred} using that component only, while using the full covariance for $\mathbf{K}$.
\end{textbox}

\subsubsection{Quasi-periodic GP models for stellar light curves}

The light curves of Sun-like stars (broadly construed) display low-amplitude quasi-periodic variations, which are caused by the rotational modulation and evolution of magnetically active regions on their surfaces. These produce quasi-periodic variations in both photometry and radial velocity observations, and GPs have in recent years become one of the most popular ways of modelling them. One reason for this is that a simple quasi-periodic covariance function (\autoref{eq:kqp}) gives rise to functions, which reproduce the light curves of rotating stars with evolving active regions remarkably well.


GP models were first applied to the simulated observations of the Sun's total irradiance variations by \citet{2012MNRAS.419.3147A}, to test a new method to simulate Radial Velocity (RV) variations based space-based photometry on solar data. The context of this work was the large numbers of candidate transiting exoplanets being discovered at the time by the \textit{CoRoT} and \textit{Kepler} space missions. Detecting the planet signals in RV was needed to confirm the planetary nature of the candidates by measuring their masses, but was being hampered by the apparent RV variations caused by active regions. Using simple geometric considerations, \citet{2012MNRAS.419.3147A} derived a simple (approximate) relationship between the flux perturbation caused by active regions, $F$, and their RV signature, which depends on both $F$ and its time-derivative $F'$. They tested this data-driven ``$FF'$ method'' on simulated photometric and RV observations of the Sun-as-a-star produced using resolved magnetograms, using a GP as a principled smoothing tool to evaluate $F$ and $F'$ from the photometry, and comparing the results to the simulated RVs. \citet{2012MNRAS.419.3147A}  tested a number of covariance functions, both aperiodic and Quasi-Periodic (QP), and found the former to be preferred in the solar case, as expected if most active regions on the Sun do not persist for much longer than its rotation period. More active and/or more rapidly rotating stars display variability that is coherent over multiple rotation periods, making QP GPs the model of choice. For example, \citet{2014MNRAS.443.2517H} used a QP kernel when applying the aforementioned $FF'$ method to the active planet-host star CoRoT-7.

Today, using GPs to model light curves containing stellar variability is standard practice. The kernel most frequently used for this purpose is the aforementioned QP kernel, whose simplicity and flexibility make it a popular choice. \citet{2018MNRAS.474.2094A} implemented a Bayesian inference framework based on this kernel to measure accurate rotation periods from \textit{Kepler} light curves. The very complex dependence of the likelihood surface on the parameters, particularly the period, makes this challenging, and careful tuning of the posterior sampling strategy is required.
These methods have since been adapted and applied to ground-based photometric surveys \citep{2020MNRAS.492.1008G}, and the shorter baseline \textit{K2} light curves \citep{2021ApJ...913...70G}.

Although the QP kernel provides a phenomenological rather that physically motivated description of the variability, some of its parameters lend themselves to a physical interpretation in terms of the star's rotation period and the evolution timescale of active regions. \citet{2022MNRAS.tmp.2007N} tested this using simulated light and RV curves based on physical star-spot models and confirm that, for moderately well-sampled datasets, the period of the GP does indeed provide a precise and accurate measure of the stellar rotation period. The same is true, albeit to a lesser extent, for the evolution timescale: the correlation between simulated and recovered values is more scattered, and breaks down when the dataset spans less than the simulated evolution timescale.

On the other hand, the QP kernel (or any covariance function of time only) does not give access to the physical properties of the spots, such as their size, latitude or contrast. The problem of inferring these properties from light curves is fundamentally ill-posed \citep{2021AJ....162..123L}, making direct inference of individual spot properties or brightness maps highly degenerate (whatever the methodology used). However, \citet{2021AJ....162..124L} derive a closed-form expression for a GP that describes the light curve of a rotating, evolving stellar surface conditioned on a given distribution of star spot sizes, contrasts, and latitudes. This can be used in a hierarchical Bayesian framework to infer the distribution in question from ensembles of light curves.

Most studies focusing on the photometric signatures of stellar activity ignore variations on short timescales, either working with binned data or using a ``jitter term'' to absorb them. This keeps the model simple and, when binning, speeds up computing time. However, when the time-sampling and precision of the observations allow it (e.g.\ for data from \textit{CHEOPS} or \textit{PLATO}), composite GP models with different terms to describe activity, granulation and stellar oscillations can be useful. For example, \citet{2020A&A...634A..75B} analysed light curves of stars observed at high cadence and high precision in the \textit{CoRoT} asteroseismology field. They showed that the parameters of planetary transits injected into these light curves were recovered more accurately when using a composite GP to model the stellar variability than one with a single term. They also tested a white noise only model, which provided less precise estimates than either GP models, albeit generally consistent with both.

\subsubsection{Stellar activity in Radial Velocities} \label{sec:RVact}
Given the remarkable instrumental precision achieved by modern RV spectrographs, stellar activity is nowadays the key factor limiting the sensitivity of RV surveys to low-amplitude and/or long-period planet signals. The effect of active regions on RV observations is two-fold. First, as dark spots rotate on the stellar surface, they distort the profile of spectral lines, removing a small contribution first from the red wing of each spectral line, then from the line core, and then from the blue wing. The second effect is more subtle: in the absence of active regions, spectral lines of Sun-like stars display a net blue-shift due to granulation: the emission from the hot, up-welling material in the granules, which is blue-shifted, dominates over that from the cooler material falling back down in the inter-granular lanes). In a facula, i.e.\ a region of enhanced magnetic flux density compared to the ``clean'' photosphere, convection is suppressed, leading to a localised reduction in this ``convective blue-shift''. Faculae have very small photometric contrast but can cover a much larger area than dark spots, so their RV signature can dominate over that of spots, especially for moderately slow rotators like the Sun with modest magnetic fields \citep{2010A&A...512A..39M}.

\begin{figure}[ht]
  \includegraphics[width=\linewidth]{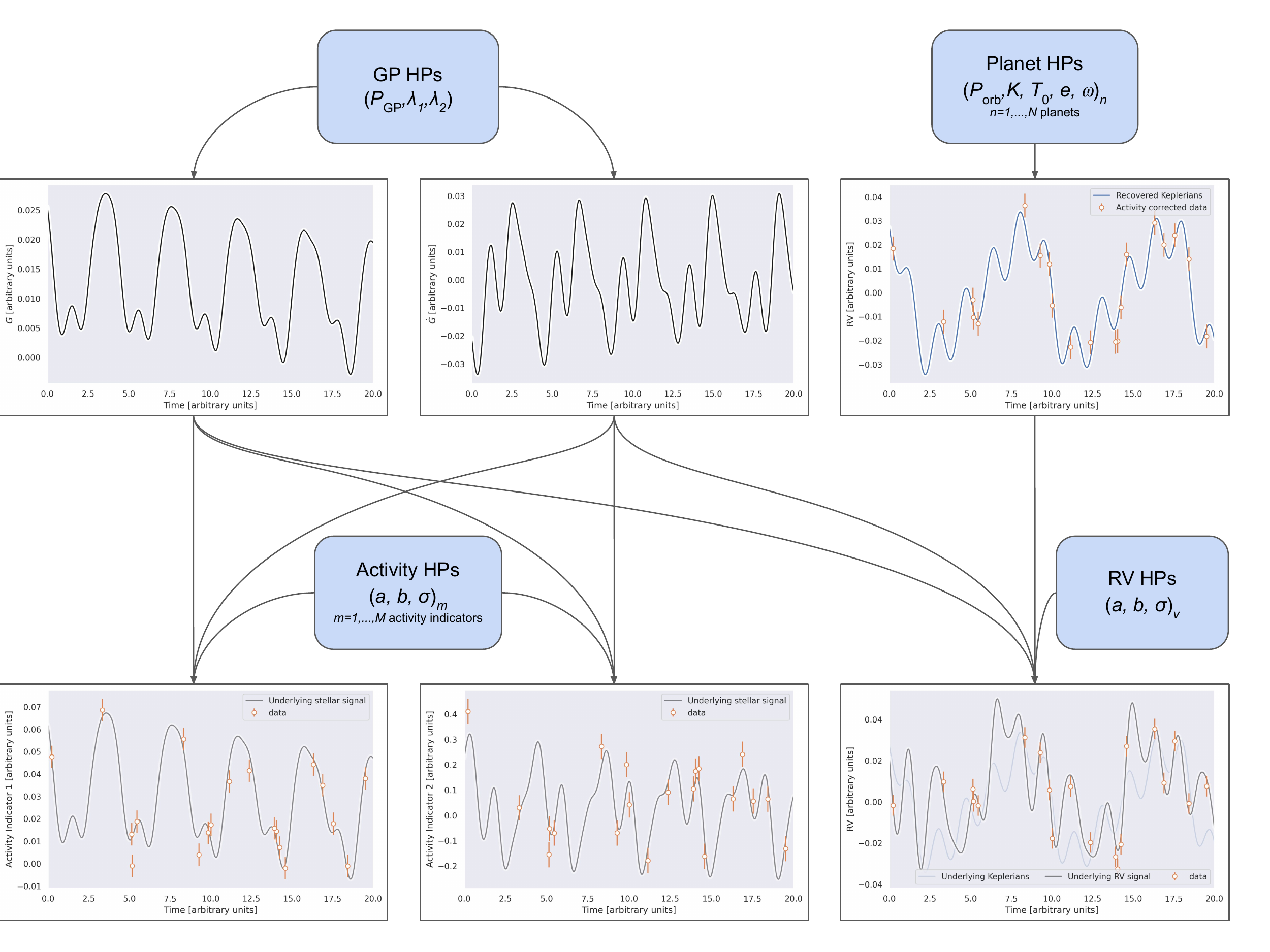}
  \caption{Schematic of the multidimensional GP framework of \citet{2015MNRAS.452.2269R} for modelling activity signals in RV data. The observed RVs (shown as orange points with error bars in the bottom right panel) are modelled alongside observations of activity indicators extracted from the spectra or line profiles (orange points with error bars in the bottom left and bottom middle panels). In all these time-series, the activity signal (shown by the black line in each panel) is modelled as a linear combination of an unobserved (latent) GP, shown in the top-left panel, and its first time-derivative, shown in the top middle panel. In addition, the Keplerian signal(s) of putative planet(s) (the sum of which is shown by the blue lines in the top right and bottom right panels) contribute to the RVs but not to the activity indicators. This hierarchical model is implemented in a Bayesian framework, sampling over all the hyperparameters listed in the blue boxes (subject to appropriate priors) to derive posteriors over the hyperparameters of interest and/or estimate the evidence for models with different numbers of planets. }
  \label{fig:RV_framework}
\end{figure}

Overall, active regions produce quasi-periodic RV variations which are frequently modelled with the same kind of QP GPs as for light curves. However, unlike transits, planetary signatures in RV occur on timescales similar to activity signals, meaning that the risk of overfitting is significant. This is partly mitigated for RV follow-up of transiting planets by strong priors on the stellar rotation period and on the period and phase of the planetary orbit, but it is a severe problem for ``blind'' RV searches. The issue is compounded by the fact that RV observations are invariably ground-based, and often have sparse time-sampling.

As previously mentioned, \citet{2012MNRAS.419.3147A} showed that an approximate relationship should exist between the photometric and RV signatures of active regions. Specifically, if $F(t)$ is the photometric signature and $F'(t)$ its time-derivative, the RV variations are expected to scale as $A F(t) F'(t) + B F^2(t)$, where the first term corresponds to spots and the second to faculae (assuming that the two are generally co-located, although the latter covers a larger area), and $A$ and $B$ are tunable free parameters. In its original form, however, this method was of limited applicability, as it requires the RV observations to be quasi-simultaneous with high-precision, tightly sampled light curves, which is the case only exceptionally (see e.g.\ \citealt{2014MNRAS.443.2517H}).

On the other hand, RV extraction pipelines routinely provide additional indicators measuring the characteristic width and asymmetry of the spectral lines, which are also affected by active regions, as well as spectral activity indicators which trace chromospheric emission in the cores of strong lines. \citet{2015MNRAS.452.2269R} developed a GP framework, schematically illustrated in \autoref{fig:RV_framework}, to model these activity indicators alongside the RVs, and to disentangle them from planetary signals. Each observed time-series is modelled as a linear combination of some latent variable $G(t)$, its time derivative $G'(t)$, white noise, and (in the case of the RVs only) one or more Keplerian signals. $G(t)$ does not have a direct physical interpretation; it loosely corresponds to $F^2$ in the $FF'$ framework of \citet{2012MNRAS.419.3147A}, so that $G'(t) \propto F(t) F'(t)$, but it is modelled as a (typically quasi-periodic) GP, which makes it possible to write down expressions for the covariance between any pair of observations from the any of the time-series included in the analysis. This results in a global covariance matrix of dimensions $(NM,NM)$, where $N$ is the number of observations and $M$ the number of activity indicators included in the analysis (including the RVs themselves). This framework has proved particularly successful for RV confirmation of transiting planets around young stars \citep{2019MNRAS.490..698B,2022MNRAS.512.3060Z}, enabling the detection of planetary signals almost 50 times smaller than the activity signals.  Open-source implementations of this framework have been published recently \citep{2022MNRAS.509..866B,2022A&A...659A.182D}, and extensions thereof are likely to play a significant role in the analysis of next-generation RV surveys targeting Earth analogues, which need to achieve even better contrast between activity and planetary signals. Very recently,  \citet{2022arXiv220506627C} proposed a GP regression network model for stellar activity signals in RV, showing that it can be used to model RVs of the Sun-as-a-star simultaneously with other activity indicators.

\subsubsection{Granulation and asteroseismology}

As discussed in \autoref{sec:physical}, asteroseismology is an area where physically motivated GP models arise quite naturally. This has motivated studies where GPs have been used to model stellar oscillations in the time-domain \citep{2017AJ....154..254G, 2018ApJ...865L..20F}, but these remain very much an exception to the much more widespread practice of transforming the data into the Fourier domain, and identifying and modelling oscillation modes therein. Using a GP in this context makes particular sense when the data are irregularly sampled or inhomogeneous, or when the GP is combined with a mean function that is much easier to specify in the time-domain (for example a planetary transit). For example, the \textit{PLATO} mission \citep{2014ExA....38..249R} aims to characterise most of the the host stars of the transiting planets it will discover up to magnitude $V \sim 11$ through asteroseismology (except for late K and M stars for which the oscillation amplitudes are too small and the frequencies too high). Current plans for the \textit{PLATO} pipeline involve either subtracting the best-fit transit model from the light curve, or cutting and interpolating over the transits, before performing the seismic analysis. However, a GP model would  enable a simultaneous analysis, which should in principle be more robust. Conversely, \citet{2020A&A...634A..75B} demonstrated the importance of 
explicitly accounting for short-term stellar variability signals, including both granulation and oscillations, to obtain robust and accurate values of the transiting planets' parameters; and compared a range of GP models with different kernels for this purpose.

\subsection{AGN variability}

The Active Galactic Nuclei (AGN) at the centre of many distant galaxies produce stochastic variability in their optical and radio emission, on timescales ranging from hours to years.
GPs can be useful descriptive models for these processes, and research in this domain has driven the development of new methods for GPR with large datasets.
In particular, as discussed in \autoref{sec:intro}, a GP was used by \citet{prh92a} to model the underlying variability of gravitationally lensed quasar 0957+561 to measure its time delay, in one of the first explicit uses of GPR in the refereed astronomical literature.

When the quasar is lensed by an intervening galaxy or galaxy cluster, multiple images can be formed, whose brightness can be monitored individually.
Measurements of the time-delays between the resulting light curves can be used to constrain the Hubble constant, $H_0$ \citep{doi:10.1146/annurev.aa.30.090192.001523}.
Early measurements of these time-delays in the decade following the discovery of the first multiply imaged quasar, the ``double quasar'' 0957$+$561, were hampered by the lack of a generative model for the light curves.
Heuristic methods based on interpolation and cross-correlation were developed, but uncertainties were difficult to derive and results obtained with different instruments and different bandpasses were not always consistent.
\citet{prh92a} derived a simple $\chi^2$ metric under the assumption that the observed light curves are shifted, noisy versions of a single sample a Gaussian process with known covariance.
Having proposed an empirical procedure to estimate the covariance function from the data, they then optimised the $\chi^2$ with respect to the time-delay between images, and applied standard methods to derive an uncertainty.
While their method relied on an ad hoc method to estimate the covariance matrix, and did not propagate the corresponding errors into the final result, it did resolve the prior discrepancy between the estimates of the time-delay from optical and radio datasets \citep{prh92b}.
Even then, the requirement to invert the covariance matrix was a barrier to the wider application of this method.
This was addressed a few years later by \citet{pr95}, who introduced a fast algorithm, to invert the covariance matrix for exponential kernels.
Updated versions of this algorithm now form the basis for fast GP solvers discussed in \autoref{sec:fast}. A closely related application domain where GPR is used to model stochastic AGN variability in order to infer a time-delay is reverberation mapping \citep[see e.g.][]{2011ApJ...735...80Z}. In this case, the time-delay of interest is between the continuum (which traces the emission from the central engine) and emission lines (which trace emission by nearby gas clouds that are photo-ionized by the continuum radiation). 

GPR is also used to study the population-level properties of AGN samples, by characterizing and interpreting the statistics of the light curve variability for large samples.
The idea is that the properties of the stochastic variations of an AGN light curve are driven by an accretion disk near the central supermassive black hole, and that this variability provides an observational probe of the small scale astrophysical processes.
The standard approach is to fit a GP model to the light curve---typically using a ``damped random walk'' kernel \citep[e.g.,][]{2010ApJ...708..927K, 2010ApJ...721.1014M, 2012ApJ...753..106M}, or the more general ``continuous-time autoregressive moving average'' (CARMA) kernel \citep[e.g.,][]{2014ApJ...788...33K, 2022ApJ...930..157Z, 2022ApJ...936..132Y}, since these permit efficient computations even with large datasets---and then to use the inferred hyperparameter values to constrain the variability amplitudes and timescales of the system \citep[e.g.,][]{2010ApJ...708..927K, 2010ApJ...721.1014M, 2012ApJ...753..106M, 2014ApJ...788...33K, 2017MNRAS.470.3027K, 2019PASP..131f3001M, 2022MNRAS.514..164S, 2022ApJ...930..157Z, 2022ApJ...936..132Y}.
These amplitudes and timescales have been shown to empirically correlate with the fundamental properties of the AGN system, such as the mass of the central black hole, Eddington ratio, and bolometric luminosity.

Driven by the success of this approach, and the scale of upcoming survey datasets from, for example, the Vera C. Rubin Observatory and the LSST survey, there has been a concerted effort within this field to develop GP methods that can scale to large datasets.
This began, perhaps, with \citet{pr95}, and the development of a fast GP solver for exponential kernels.
While widely used, this kernel is very restrictive, and more flexible models are required to capture the full range of variability observed in AGN light curves.
To this end, \citet{2014ApJ...788...33K} developed a scalable GP solver for the CARMA kernel, which uses a Kalman filtering approach to achieve linear scaling of the computational cost with the number of data points, which we discuss further in \autoref{sec:fast}.

\subsection{Compact objects, gravitational waves, and transients}

GPs have also been used for the analysis of a broad range of other astrophysical objects, including pulsars, gravitational wave sources, and other transients.
In this section, we highlight some of these remaining applications, in somewhat less detail than the other applications since we have less familiarity with these topics.
In many of these domains, it is standard practice to work in the Fourier domain by computing the fast Fourier transform (FFT) of the data, and then assume that the noise process is uncorrelated between frequency bins \citep[e.g.,][]{2017LRR....20....2R}.
In other words, the data covariance matrix in the Fourier domain is assumed to be diagonal, and this significantly reduces the computational cost of likelihood evaluations compared to a GP model with a dense covariance matrix.
This is a reasonable assumption for datasets with long observational baselines relative to the signals of interest, and stationary noise processes \citep[e.g.,][]{UNSER1984231}.
While these requirements are not met by many of the analyses that we have discussed, large datasets from pulsar timing surveys and gravitational wave interferometers do often satisfy these constraints, or can be transformed in such a way that they do.

\subsubsection{Pulsar timing} GPs are widely used in the analysis of pulsar observations, as a model of timing residuals, pulse shape variations, and the stochastic signals of interest.
In early work, \citet{2014PhRvD..90j4012V} describe a comprehensive set of GP models, and computational techniques for their implementation, for pulsar timing studies aimed at detecting gravitational waves.
These models include the effects of timing noise, dispersion-measure variations, and the stochastic gravitational wave background.
Such GP models are now routinely used for pulsar timing studies \citep[e.g.][]{2019MNRAS.489.3810P, 2020MNRAS.498.6044C, 2022MNRAS.510.4873A}.

GPs have also been used to model pulse shape variations in decades-long pulsar observations \citep{2016MNRAS.456.1374B,2019MNRAS.488.5702B,2022MNRAS.513.5861S}.
In this case, the GP is used to model the pulses themselves, and to track long-term changes in pulse periodicity and shape.
This is done by evaluating (analytically) the first and second derivatives of the fitted GP and using those to find extrema and inflection points of the pulse profile, along with the associated uncertainties.
Finally, \citet{2021MNRAS.508.4249P} used GPs to characterise the off-pulse noise background in observations of pulsars with large duty cycles and no obvious flat baseline.

\subsubsection{Gravitational waves} Within the LIGO--Virgo project, GP models are used to model the detector noise \citep[e.g.][]{2020CQGra..37e5002A}, and waveform modelling uncertainties \citep{2016PhRvD..93f4001M}.
For these datasets, the power spectrum has been empirically calibrated, and it includes non-trivial structure at all frequencies.
As discussed above, under some generally reasonable conditions, this noise process diagonalizes in the Fourier domain, and this empirically calibrated model can be used to compute the likelihood of a signal model also represented in the Fourier domain.
This type of model becomes significantly more complicated for ring-down analyses where edge effects invalidate the assumption that the data are periodic \citep{2021arXiv210705609I}.

\subsubsection{Transient classification} In the context of transient light curve classification, GPs have been widely used as a tool for interpolating sparsely-sampled and noisy photometric time series onto uniformly sampled grids that are more amenable to standard machine learning classification methods \citep[e.g.][]{2016ApJS..225...31L, 2019AJ....158..257B, 2019MNRAS.489.3591P, 2020ApJ...905...94V}.
These pipelines differ in how they treat the covariances between photometric bands, with most learning a correlation length in effective wavelength.
Recently, \citet{2021ApJS..255...24V} proposed a wavelength covariance matrix that uses a distance metric defined by the throughput overlap between each pair of filters.
So far in this domain, GPs have primarily been used in a pre-processing step to produce data correctly structured for standard machine learning methods, but recent developments that integrate GPs with deep learning frameworks could permit these GPs to be embedded within the larger classification pipeline.

\subsubsection{Quasi-periodic oscillations}

One very active area of research where GPs have not been widely applied to date, is the characterisation of Quasi-Periodic Oscillations (QPOs). QPOs are seen in astrophysical transients right across the electromagnetic spectrum, including Gamma-ray bursts, X-ray binaries, flares in Sun-like stars and in magnetised pulsars, and fast radio bursts. Most QPOs are believed to arise from the interaction between one or more compact objects and an accretion disk, but a detailed understanding of the origin of most classes of QPOs remains elusive. Their phenomenology is typically studied in the time domain, by fitting the power spectrum of individual, contiguous blocks of observations with empirical models, most frequently a sum of Lorentzian functions \citep[see][and references therein]{ingram_motta_QPOs}. Very recently, two studies \citep{2021ApJ...907..105Y, 2022ApJ...936...17H} used GPs to model QPOs directly in the time-domain, with encouraging results both for the detection of QPOs in the presence of correlated noise and the characterisation of their properties such as frequency.

\section{CHALLENGES, PITFALLS, AND SOLUTIONS}
\label{sec:challenges}

So far, we have primarily discussed the benefits of GP models for astrophysics, but in this section we aim to highlight some challenges faced when using GPs in practice.
The most significant challenge that has so far limited broader adoption GPs for astrophysics is their computational cost and scaling, but as we will see below, several techniques have been developed recently that overcome this limitation, in particular for one dimensional datasets that are common in time domain astronomy.
We also briefly discuss some contexts where GP models may overfit, or be otherwise misspecified for specific problems.

\subsection{Scalable GP inference}
\label{sec:fast}

\begin{figure}[ht]
  \centering
  \script{scaling.ipynb}
  \includegraphics[width=0.7\linewidth]{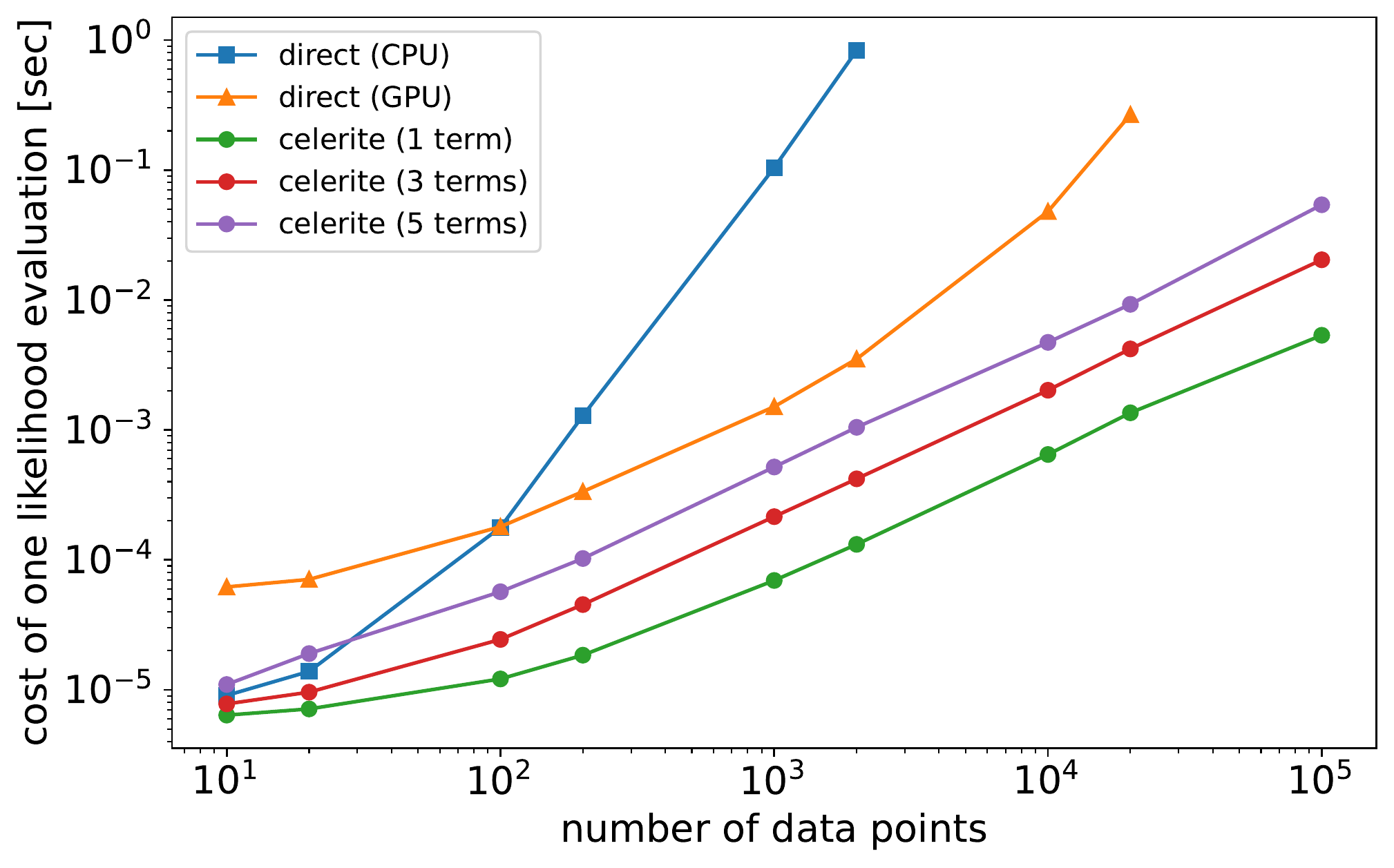}
  \caption{Computational cost of a single evaluation of the GP likelihood function (\autoref{eq:gp_like}) versus the size of the dataset using different algorithms.
  In each case, the kernel function is a one-dimensional Mat\'ern-3/2 with randomly generated data.
  The top two lines labelled ``direct (CPU)'' and ``direct (GPU)'' show the cost of evaluating \autoref{eq:gp_like} directly using the Cholesky factorization of $\mathbf{K}$ provided by a tuned and optimised linear algebra library.
  The CPU solver was run using a single-threaded implementation, and the GPU version was executed on an \textsf{A100 NVIDIA} GPU.
  In both cases, the asymptotic scaling of the algorithm is approximately $\mathcal{O}(N^3)$ for $N$ data points.
  The three lines labelled ``celerite ($J$ terms)'' show the performance of the \project{celerite} algorithm \citep[][as implemented by the \project{tinygp} package]{celerite} for a sum of $J$ Mat\'ern-3/2 kernels, since the computational cost of that method scales with the ``complexity'' of the kernel.
  The asymptotic performance of each of these curves is linear in the number of data points, $\mathcal{O}(N)$.}
  \label{fig:scaling}
\end{figure}

The core computations required for GPR are the likelihood function (\autoref{eq:gp_like}), and predictive distributions (\autoref{eq:pred}).
All of these computations require the solution to a linear system, whose dimension scales with the number of data points $N$.
While this can be solved using optimised linear algebra libraries, possibly including acceleration with a GPU or other highly parallel hardware, the raw computational cost of evaluating a GP model typically scales as the cube of the number of data points $\mathcal{O}(N^3)$, as shown in \autoref{fig:scaling}.
This makes it intractable to use GPs for large datasets, practically restricting their use to datasets with less than a few hundred data points.
General approaches have been developed for scaling GP models to larger datasets, and there are some methods that have seen significant impact within the astronomical literature.

Scalable methods for GPR typically fall into two broad---but sometimes conceptually overlapping---categories: \emph{(1)} approximate, or \emph{(2)} exact on a restricted function space.
Approximate methods typically rely on algorithms from computational linear algebra to construct sparse, low-rank, or otherwise structured approximations to the covariance matrix that can be used to approximately solve the required systems to within a target numerical accuracy.
On the other hand, there exist scalable exact methods that are restricted to a particular class of covariance functions, or data with a specific structure (for example, one-dimensional evenly sampled data).
Approximate methods \citep[e.g.,][]{inducing, george, kissgp, 2015arXiv151101870W} have been widely used in the astrophysics literature thanks, in part, to the existence of publicly-available software (e.g.\ \project{GPy}\footnote{\url{https://github.com/SheffieldML/GPy}}; \project{george}, \citealt{george}; \project{GPyTorch}, \citealt{gpytorch}).
However, within time-domain astronomy, scalable exact methods have been more widely used, since time series data often has more amenable structure.

One simple example of a scalable exact method is applicable to one-dimensional evenly sampled data, where the covariance matrix has ``Toeplitz'' structure \citep[e.g.,][]{toeplitz}, which permits more efficient computations using a fast Fourier transform.
The use of Toeplitz structure for scalable Gaussian noise modelling has primarily been applied for gravitational wave data analysis \citep[e.g.,][]{2020PhRvR...2d3298T, 2021arXiv210705609I}, however, most datasets in astrophysics are not evenly sampled, limiting the applicability of this method.

The most widely-used algorithms for scalable exact GPR in time domain astrophysics are based---often not explicitly---on stochastic differential equations \citep[SDEs;][]{Sarkka:2019}, and typically require that the data be ``sortable''.
As an early example in the astrophysics literature, \citet{pr95} presented an algorithm for GPR with linear scaling with the size of the dataset, when using an exponential kernel with unevenly sampled time series data.
This method was later generalized to a wider range of covariance functions \citep{ambikasaran2015generalized, celerite, 2020AJ....160..240G, 2020A&A...638A..95D, 2022A&A...659A.182D}.
The real-world performance of these generalized methods (as implemented by the \project{tinygp} package) is shown in \autoref{fig:scaling} by the curves labelled ``celerite'', and indexed by the number of ``terms'' (in this case Mat\'ern-3/2 functions) in the kernel \citep[see][for a more detailed discussion]{celerite}.
The \project{celerite}\footnote{\url{https://github.com/dfm/celerite}} and \project{S+LEAF}\footnote{\url{https://obswww.unige.ch/~delisle/spleaf/doc}} implementations of these methods have been widely-used for a wide range of applications, including: measuring stellar rotation periods \citep[e.g.,][]{2020MNRAS.492.1008G, 2021ApJ...913...70G, 2022AJ....164..115N}; characterizing AGN light curves \citep[e.g.,][]{2022ApJ...936..132Y}; and modelling the light curves of transiting planets \citep[e.g.,][]{2019AJ....158...32K} and eclipsing binaries \citep[e.g.,][]{2020ApJS..250...34C}.
A qualitatively different, and independently developed approach to implementing these models uses a state-space representation for the underlying SDE to enable scalable inference using a Kalman filter \citep{2014ApJ...788...33K,  2021RNAAS...5..107J, 2022arXiv220709327M}.
Of particular interest, while the core algorithms discussed here are inherently serial and cannot benefit from parallel hardware, this filtering approach to GPR has recently been formally parallelized \citep{sarkka2020temporal}.

\subsection{Overfitting}

In our experience, a common concern that astrophysicists have when working with GPs is that these flexible models may overfit the observations, and decrease the significance of their results, especially when a GP is used as an effective model for noise sources.
While this is not an unjustified concern, we find that the risk is generally overestimated.
As discussed in \autoref{sec:gp-lsq-link}, the determinant in \autoref{eq:gp_like} penalises GP models with ``unspecific'' hyperparameters, instead preferring to explain as much of the data as possible with the mean model.
In other (but equally informal) terms, this means that a GP's modelling capacity will only be used to describe residuals that are not captured within the mean model's support.

All this being said, there do exist cases where the use of a GP can lead to overfitting. One such example is RV planet searches when the ephemeris of the planet(s) is unknown (i.e.\ excluding transiting planets). Unless the time-sampling of the data is truly excellent, the hyperparameters of the GP activity model are poorly constrained, and the highly multi-modal nature of the posterior surface means that standard Bayesian inference and model comparison techniques can be misleading. One example of such a situation is the short-period, Earth-mass planet reported by \citet{2012Natur.491..207D} around $\alpha$\,Centauri\,B, which was shown by \citet{2016MNRAS.456L...6R} to be an artefact of the time-sampling: a small peak in the window function of the observations was serendipitously ``boosted'' by the complex model necessary to account for the star's activity signal. This is a concern with any flexible model applied to a sparse dataset, but it can be less obvious for GPs, which can be very flexible with comparatively few hyperparameters. Useful strategies to deal with this kind of problem include cross-validation (as discussed in \autoref{sec:model_selection}) and performing exhaustive injection-recovery tests to build (or undermine, as the case may be!) confidence in one's results. 

\subsection{Model misspecification}

Like any other model for data, GPs are subject to the risk of model misspecification.
A particularly common example of model misspecification in astronomy is the use of a Gaussian observation model when there are ``outliers'' in the dataset with extremely large residuals.
These outliers can be caused by a wide range of phenomena (e.g., instrumental artefacts, cosmic rays, etc.), but cannot be properly modelled as generated by Gaussian noise.
Model misspecification when faced with outliers does not uniquely affect GP models, and approaches have been developed to handle these problems in general.
In the absence of correlated noise, a probabilistic approach that is commonly used in astrophysics is to build a two component mixture model for the observations \citep[e.g.,][]{1997upa..conf...49P, 2010arXiv1008.4686H}, which can be used to infer which data points are outliers.
This method does not generalize to GP models since, when there is correlated noise, the observation model is no longer independent over data points:
every observation ``cares'' about whether or not every other measurement is an outlier, since we must compute the covariance between every pair of points.

In the astrophysics literature, the most commonly used approach to handling outliers in GP models is to perform an initial iterative ``sigma clipping'' procedure to identify and remove outliers \citep[e.g.,][]{2019ApJ...885L..12D}.
There are numerous sigma clipping procedures used in the literature, but a common approach is to: \emph{(1)} fit the GP model (including the mean model) to the full dataset using a numerical maximum likelihood procedure (\autoref{sec:inference}); \emph{(2)} compute the predictive distribution (\autoref{sec:pred}) at the observed times; \emph{(3)} remove data points with significantly low likelihood under this predictive distribution; \emph{(4)} iterate from step (1) with this clipped dataset until no more outliers are removed.
This procedure can work well when the outliers are rare and extreme, but may not be appropriate for all use cases.

More generally, outside of astronomy, there has been some work on developing a framework for Student's-$t$ processes as generalisations of GPs, at no additional computational cost \citep[e.g.,][]{shah2014student, Tracey_2018}.
A Student's-$t$ distribution can be used to model processes with heavier tails than a GP, so a Student's-$t$ process may be more robust to model misspecification.
However, the Student's-$t$ process in and of itself is not a good model for dataset with outliers because the underlying functions are still required to be smooth.
Instead, a more promising approach might be to use a GP model for the ``true'' function, combined with a per-observation independent Student's-$t$ noise model \citep[e.g.,][]{NIPS2009_13fe9d84}.
The practical application of this method is non-trivial and potentially extremely computationally expensive, since the marginal likelihood of such a model can no longer be evaluated in closed form.

Given the current state of the field, our current recommendation for assessing and mitigating model misspecification is to use the methods discussed in \autoref{sec:model_selection} to determine the suitability of a GP for a specific application.
In the case where there are significant outliers in the dataset, the most mature and commonly used approach is to perform iteratively sigma clip the data as a pre-processing step.
As an alternative, \citet{2015ApJ...812..128C} propose a method for handling non-stationary and structured model misspecification in the context of stellar spectral modelling, by designing a basis of ``local'' kernel functions tuned to capture line shape residuals.
Such domain-specific methods can be developed and validated using the methods discussed in \autoref{sec:model_selection}.

\section{OPEN-SOURCE GAUSSIAN PROCESS SOFTWARE}
\label{sec:open}

It is reasonably straightforward to implement a simple GP model in code, and within astrophysics it has been common for authors to implement custom GPs for their analysis.
However, things get significantly more complicated when implementing the scalable or approximate methods discussed in the previous section.
Similarly, it can be tedious to experiment with different kernel functions and inference methods without building a non-trivial modelling infrastructure, something that has typically been ad-hoc in astrophysics research.

Luckily, driven by the immense popularity of GPR in machine learning, the physical sciences, and other fields, there are a plethora of open source tools that have been developed to simplify this process for a wide range of applications.
Many of these tools are designed for scalability, flexibility, and ease of use.
In this section, we describe some popular libraries in this space, while cautioning the reader that this is not a comprehensive list, and this domain changes quickly, so the discussion may become outdated more quickly than the rest of this document.
Most of our discussion will focus on tools implemented in the \project{Python} programming language since it is---at the time of writing---the most popular language, both in astronomy and GP modelling, but there are tools available in all other popular languages.

Alongside this review, we have released all the source code to generate the figures in this document using the \project{tinygp} library\footnote{\url{https://tinygp.readthedocs.io}} \citep{tinygp}, which is a relatively new open source \project{Python} library for GP modelling.
While this library has not yet been widely used in the astrophysics literature, we believe\footnote{Note that the development of the \project{tinygp} library is led by D.\ Foreman-Mackey, an author of this review.} that it is a good choice for many applications.
\project{tinygp} uses the \project{JAX}\footnote{\url{https://jax.readthedocs.io}} library as its computational backend, giving it good computational performance (including support for GPU hardware) with a pure-\project{Python} user interface.
\project{JAX} also provides support for automatic differentiation \citep[e.g.,][]{autodiff}, which can be used to accelerate all the inference schemes described in Sections~\ref{sec:inference} and~\ref{sec:model_selection}.

Perhaps the most widely used \project{Python} library for GP inference in astrophysics is \project{george}\footnote{\url{https://george.readthedocs.io}} \citep{george}, which is a relatively simple library that implements most of the standard kernel functions described in \autoref{sec:kernels}, and an approximate solver that can be used to scale to moderately large ($N \sim 1000$) datasets.
More recently, the scalable methods implemented by \project{celerite}\footnote{\url{https://github.com/dfm/celerite}} \citep{celerite} and \project{S+LEAF}\footnote{\url{https://obswww.unige.ch/~delisle/spleaf/doc}} \citep{2020A&A...638A..95D, 2022A&A...659A.182D} have become popular for time domain GP models in astrophysics.
These libraries require one-dimensional---or appropriately structured \citep{2020AJ....160..240G, 2022A&A...659A.182D}---input data, and a specific set of kernel functions, although these kernels form a complete \citep{JMLR:v22:21-0072} and physically interpretable \citep{celerite} basis, and it has recently been demonstrated that the Mat\'ern class of kernels can be exactly represented \citep{2021RNAAS...5..107J}.

Outside of astrophysics, the \project{GPy}\footnote{\url{https://github.com/SheffieldML/GPy}}, \project{GPyTorch}\footnote{\url{https://gpytorch.ai}} \citep{gpytorch}, and \project{GPFlow}\footnote{\url{https://www.gpflow.org}} \citep{GPflow2017, GPflow2020multioutput} projects provide mature and widely used GP implementations.
\project{GPyTorch}, in particular, is designed for scaling GP modelling to large datasets, using approximate methods \citep[e.g.,][]{kissgp, 2015arXiv151101870W} that do not have the same restrictions as \project{celerite} and \project{S+LEAF}.
\project{GPyTorch} has been used for a variety of applications in astrophysics \citep[e.g.,][]{2022A&A...658A.166D, 2022MNRAS.511.5597L}, but not yet as widely as the other libraries discussed so far, perhaps because it does not integrate into the astronomical workflow as easily as the projects that were designed with astrophysicists in mind.

\section{CONCLUSIONS}
\label{sec:concl}

Gaussian Process Regression has become a key element in the toolkit of modern astronomers, particularly those working with time-domain datasets. It offers a principled approach to modelling stochastic signals and nuisances, particularly correlated noise, whose importance has grown significantly in the past decade with the development of major precision time-domain monitoring surveys (e.g.\ planetary transit surveys). GPR is explicitly 
rooted in probability theory, making it readily interpretable and enabling astronomers to quantify their confidence (or lack thereof) in the results of their analysis. Despite the simplicity of the underlying mathematical framework, even basic GP models are extremely flexible, and a small handful of commonly used ``building blocks'' have been used successfully to explain the observed variability of astronomical objects on all scales. On the other hand, this flexibility comes at a price: a ``good fit'' with a particular model is no longer a fail-safe indication that this particular model is ``the correct one'', so users must proceed with extra caution. 

\subsection{Summary of the review}

After a very brief historical introduction, we started this review by introducing two simulated examples, which demonstrate the two main classes of use for GPR in astronomy, namely to account for a nuisance signal (in the case of a planetary transit) or to model a stochastic signal of interest (in the quasar time-delay example). The source code for these and other worked examples in the review is available alongside this review in the form of \project{Jupyter} notebooks that can be accessed by clicking on the icon next to each figure caption, and interested users are encouraged to use these to further their understanding of the methodology and potentially adapt the code to their own datasets. Next, we gave a succinct introduction to the key equations underlying GPR regression, and sketched out a typical workflow for applying GPR to a new dataset. Like any data analysis method, GPR requires the user to make a number of choices, the most important of which is the choice of covariance function. We listed some of the most widely used covariance functions, outlined how to construct more complex ones, and gave practical advice on how to assess the validity of one's modelling choices. These strategies include standard Bayesian model comparison but also cross-validation, which can be more appropriate in cases where predictive power is important. 

We then reviewed the growing literature on applications of GPR to time-domain datasets, from exoplanet detection and characterisation, where GPs are mostly used to account for instrumental or stellar nuisance signals, to the characterisation of stellar variability itself, where the GP is now used to model a stochastic signal for its own sake. These examples are drawn from the exoplanet and stellar astrophysics literature, which we are most familiar with, but we also attempted to give an overview of recent applications of GPR to two other classes of astronomical objects where it is increasingly important, namely AGN and compact objects. Altogether, these applications include a very diverse array of GP models with varying levels of complexity, including multidimensional inputs and outputs, derivative and integral observations, and applications where the GP model is physically motivated versus cases where it is merely an effective model. We noted how GPR can be used to tease apart the deterministic and stochastic components of a model, but also different components of a stochastic process. 

We discussed the challenges and pitfalls of GPR, noting recent algorithmic and technical developments that have brought down their computational cost, once prohibitive for large datasets, and outlining some strategies to limit the risk of overfitting. Nonetheless, the size of the dataset remains an important consideration when deciding whether or not to use GPR for a particular problem. If too small, learning the kernel hyperparameters and flagging outliers can be difficult, while computing time can still be problematic for certain types of kernels and/or computations on very large datasets. Finally, we reviewed the ecosystem of free software resources for GPR, which make it easy to test a GP model for a new problem in just a few lines of code, and to implement a computationally efficient version suitable for a large dataset with minimal additional effort. 

\subsection{Future perspectives}

In certain application domains, such as the analysis of photometric and radial velocity time-series from exoplanet surveys, GPR is already used fairly widely. Its usefulness as an effective model for astrophysical or instrumental nuisance signals is well established, and it is seen as a fairly standard element of the astronomer's data analysis toolkit. In these existing applications, we foresee three main routes for further progress in the near future. One is the development of more physically motivated (as opposed to merely effective) GP models for stellar variability, so that GPR can be used to characterise and understand these stellar signals for their own sake. The other is a more systematic use of GPR as part of the de-trending and detection pipelines for exoplanet transit and RV surveys. Computational cost considerations would have made this impractical just a few years ago, but this obstacle has been lifted thanks to the recent development of scalable GP models. Finally, as JWST begins to deliver much exoplanet atmosphere observations with much higher signal-to-noise (including, importantly, more precise and relevant housekeeping parameters), a new generation of systematics models will no doubt be developed to get the most out of these observations, and it seems likely that GPR would have a role to play in that.

In other applications, such as pulsars, QPOs and astrophysical transients, GPR remains more of a niche pursuit so far, but promising early results should motivate further work in the next few years. The next decade will see a vast increase in the scale of the relevant datasets, with major surveys such as the Vera Rubin Observatory's Legacy Survey of Space and Time (LSST) or the Square Kilometre Array (SKA) routinely delivering tens of TB per night. Even with the recent development of scalable GP models, it is unclear whether GPR cam be useful as a ``first contact'' analysis tool for these very large surveys, but it has clear potential for the analysis of specific subsets of data.

One area of active development concerns the problem of non-Gaussianity. While the assumption of Gaussianity implicit to GPR is often good enough in practice, it is almost never strictly correct, and this can become problematic when extreme precision is required on the model hyperparameters, or when the data contain (a relatively small proportion of) significant outliers. Stochastic process models based on distributions with higher kurtosis than the Gaussian, for example Student's-$t$ processes are an interesting alternative, though they have not been extensively used in astronomy.

Another area where GPR could become more widely used in future is for intelligent observation planning: if a set of variable objects is being monitored as part of a survey, using a GP to model existing observations would enable the observer to make predictions (including uncertainties) for future observations, and therefore to prioritise among the target list in real time according to some metric of choice. 

\subsubsection*{Parting remarks} GPs are an extremely powerful and general data analysis method, which explains their rapidly growing popularity in a wide range of application domains, including time-domain astronomy. However, their very flexibility, together with their built-in property of automatic marginalisation over random functions, can make them appear arcane, occasionally leading to lukewarm reception from the wider community. We hope this review will have de-mystified the subject somewhat, and provided our readers with enough information to make informed decisions about when and how to use GPs for their datasets.

\section*{DISCLOSURE STATEMENT}
The authors are not aware of any affiliations, memberships, funding, or financial holdings that
might be perceived as affecting the objectivity of this review.

\section*{ACKNOWLEDGMENTS}
S.\ A.\ wishes to acknowledge support from the European Research Council (ERC) under the 
European Union's Horizon 2020 
research and innovation programme (Grant agreement No. 865624), and useful discussions with Steve Roberts, Mike Osborne, Oscar Barrag{\'a}n, Vinesh Rajpaul, and Aris Karastergiou.

D.\ F.\ M.\ acknowledges useful conversations with
Eric Agol,
Will Farr,
Alex Gagliano,
Tyler Gordon,
and Maximiliano Isi.

The authors would like to thank the community members who sent feedback on the public draft of this review:
Eric Agol,
Nestor Espinoza,
Christopher Kochanek,
Konstantin Malanchev,
Maria Pruzhinskaya,
Steve Roberts,
Fergus Simpson,
Joshua Winn,
Dahai Yan,
and
Ying Zu.

This research has made use of NASA's Astrophysics Data System Bibliographic Services.

%

\bibliographystyle{ar-style2}
\bibliography{bib}

\section*{RELATED RESOURCES}

SA gave a set of lectures on GPR for time-domain astronomy as part of the 2021 \emph{Saas Fee Advanced Course in Astrophysics}, for which PDF lecture notes are available at \url{shorturl.at/hrstz}, along with \project{Python} exercises in the form of \project{Jupyter} notebooks.

\end{document}